# Variational quantum algorithm based on Lagrange polynomial encoding to solve differential equations

Josephine Hunout,[*] Sylvain Laizet, and Lorenzo Iannucci

*Department of Aeronautics, Imperial College of London, London SW7 2AZ, United Kingdom*

Differential equations (DEs) serve as the cornerstone for a wide range of scientific endeavors, their solutions weaving through the core of diverse fields such as structural engineering, fluid dynamics, and financial modeling. DEs are notoriously hard to solve, due to their intricate nature, and finding solutions to DEs often exceeds the capabilities of traditional computational approaches. Recent advances in quantum computing have triggered a growing interest from researchers for the design of quantum algorithms for solving DEs. In this work, we introduce two different architectures of a novel variational quantum algorithm (VQA) with Lagrange polynomial encoding in combination with derivative quantum circuits using the Hadamard test differentiation to approximate the solution of DEs. To demonstrate the potential of our new VQA, two well-known ordinary differential equations are used: the damped mass-spring system from a given initial condition and the Poisson equation for periodic, Dirichlet, and Neumann boundary conditions. It is shown that the proposed new VQA has a reduced gate complexity compared to previous variational quantum algorithms, for a similar or better quality of the solution.

## I. INTRODUCTION

Solving equations and particularly differential equations (DEs) is a key element in various fields of science. In aerospace engineering, they are essential to model and understand the dynamic behavior of physical systems in structural analysis or fluid dynamics. In classical computation, DEs can be solved using differentiation methods such as finite-difference, finite-element and -volume methods, or using spectral methods like polynomial approximations. With these methods, reaching a high accuracy often requires a fine spatial discretization or a large basis set, and in both cases it can lead to a significant computational cost. In engineering applications, despite DEs typically having low dimensions from one dimension (1D) to three dimensions (3D), the need for a very fine mesh in all spatial directions is common, which can result in high memory and computational requirements.

Quantum computing is a transformative new paradigm which takes advantage of the quantum phenomena seen at the microscopic physical scale. While significantly more challenging to design, quantum computers can run specialized algorithms that can scale better than their classical counterparts, sometimes exponentially faster. It is therefore natural to investigate the potential of quantum computers for solving DEs. Quantum algorithms for tackling DEs fall into two distinct groups: fully quantum algorithms, which utilize quantum circuits to manipulate the quantum state in accordance with the relevant DE, and hybrid quantum-classical algorithms, in which a quantum computer plays a role in a broader classical computing process.

Fully quantum approaches do offer certain advantages in solving linear DEs because quantum operators act linearly on quantum states. However, the evaluation of the performance and quantum advantage of these algorithms is still unclear, as these metrics highly depend on the complexity of the problem and it is often challenging to test the algorithms on real quantum machines. Fully quantum algorithms face inherent limitations due to the necessity of encoding and retrieving extensive classical data in a quantum superposition. The advancement of this technology, known as quantum random access memory (qRAM) [1], is crucial for a significant application of such algorithms. The current state of quantum hardware and the challenges associated with fault-tolerant error correction make the implementation of a reliable and large-scale qRAM unfeasible at present [2]. Although the quick evolution of quantum computers and the symbolic limit of 1000 qubits reached with the release of the IBM Condor quantum computer [3], a large number of physical qubits do not guarantee a large number of logical qubits due to actual error-correction techniques [4].

For these reasons, hybrid quantum-classical algorithms have been developed in recent years. In particular, variational quantum algorithms (VQAs) are notable for their hybrid quantum-classical approach, making them suitable for implementation on near-term quantum devices that are not yet fault tolerant or fully error corrected. Such systems are often referred to as noisy intermediate-scale quantum (NISQ) devices. The "variational" aspect in VQAs refers to the use of variational principles, which are a mathematical approach used to find approximations to the lowest-energy states of a system. This is akin to finding the most stable configuration of a physical system. In the context of quantum computing, these principles are used to find the state of a quantum system that minimizes a certain objective function, which is often related to the problem being solved. The hybrid nature of VQAs, leveraging both quantum and classical computing resources, makes them versatile and adaptable to various types of problems and quantum hardware. This adaptability positions VQAs as a promising approach for exploring the potential of quantum computing in the near term.

---

[*]Contact author: j.hunout21@imperial.ac.uk

Among the earliest algorithms that demonstrated the potential of quantum computers to solve specific problems more efficiently than classical counterparts were the Shor algorithm [5] and the Grover algorithm [6]. Both of these algorithms were foundational in quantum computing and highlighted the potential of quantum solvers to address specific computational challenges more efficiently than classical methods. They spurred significant interest in research and development in quantum computing, leading to the exploration of various quantum algorithms for solving a wide range of problems, including those involving DEs. Many of the first quantum solvers developed in the 2000s were encoded by vectors of amplitudes and built on the quantum amplitude amplification and the quantum amplitude estimation algorithms (QAAA and QAEA) [7]. Hence, the first quantum solvers for DEs used differentiation methods over discretized spaces of variables. In 2006, the Kacewicz quantum algorithm [8] was developed to solve initial value problems (IVP), i.e., ordinary differential equations (ODEs) with initial conditions. In this quantum algorithm, the QAAA and QAEA [7] are used to estimate the mean value of the ODE over numerous sets of points.

Later, the Kacewicz quantum algorithm has been improved to be applied to DEs and simulate the solution of the steady state of the Navier-Stokes equations [9], the Burgers' equation [10], and the heat equation [11].

Another important quantum solver for DEs is the famous HHL Algorithm, a quantum solver developed in 2009 [12] which has demonstrated an exponential speedup over classical solvers in certain cases. Since then, several solvers for linear DEs have been proposed, with algorithms based on the finite-difference method [13–16], the finite-element method [17,18], high-order methods [19,20], or spectral methods [21]. Quantum solvers have also been used to solve nonlinear ODEs and partial differential equations (PDEs), often by applying linearization techniques [22–24].

Regarding VQAs specifically, they have been studied for many applications, including equation solvers, and have shown potential near-term quantum advantages. The aim of VQAs, often referred as quantum neural networks, is to classically train one or multiple parametrized quantum circuits to solve a given problem. For example, a well-known VQA is the variational quantum eigensolver [25–27] designed to find the ground-state energy of a given system, which is particularly used in quantum chemistry and materials science. VQAs have already been designed and tested to solve PDEs [28–30] or ODEs [31].

One of the most challenging aspects of VQAs is their trainability. Training variational circuits using gradient-based optimizers often prove daunting, given the exponential disappearance of the gradient with the system's size, referred as the barren-plateaus phenomenon. Nevertheless, strategic approaches, such as wise choices of observables, gradient determination techniques, and limiting the depth of the system, offer potential solutions to overcome the barren-plateau obstacle [32,33].

In this work, two architectures for a novel VQA based on Lagrange polynomial encoding are presented and compared to existing VQAs, specifically the algorithms proposed by Kyriienko *et al.* [29] (modified so that it can deal with high-order DEs) and Sato *et al.* [31]. The Kyriienko *et al.* algorithm, which has been extensively tested with first-order DEs with convincing results, encodes functions using Chebyshev polynomials, focusing on achieving high accuracy with polynomial-based feature maps. The Sato *et al.* algorithm, which has been successfully tested for the Poisson equation, is a discretized approach that approximates the solution to differential equations. In this work, the VQA is based on the Lagrange polynomials encoding. The proposed strategy leverages differentiable quantum circuits employing Lagrange polynomials and the Hadamard test differentiation method.

The two architectures proposed for the proposed new VQA differ in terms of the number of qubits required. One has a more condensed design, requiring fewer qubits but a higher number of basic gates, while the other is a larger architecture requiring more qubits but fewer gates. Despite these structural differences, both architectures produce the same theoretical output in a noise-free simulation environment. This study evaluates these architectures in terms of solution quality and algorithmic complexity, focusing on the number of gates and circuits involved. To assess the potential of the proposed VQA architectures, this study focuses on solving two differential equations (DEs): the damped mass-spring system with given initial condition and the Poisson equation under periodic, Dirichlet, and Neumann boundary conditions. All VQAs are simulated in a noise-free environment to provide a fair comparison.

The paper is organized as follows: First, a background section defines VQAs and highlights key elements such as quantum circuit differentiation methods, and existing VQAs for DEs, which serve as a basis for comparison with the VQA architectures proposed in this article. Then, methods used for the development of the novel VQA are presented, and finally results for both applications are discussed. The paper is ended with a conclusion.

## II. BACKGROUND

### A. Variational quantum algorithms

VQAs have been developed to obtain a quantum advantage using NISQ devices, i.e., with limited qubit resources, connectivity, and circuit depth, setting them apart from fault-tolerant quantum algorithms. The objective of a VQA is to ally the strengths of both classical and quantum computing paradigms, by training a parametrized quantum circuit with a classical optimizer. While VQAs have been explored across diverse applications, they may feature different algorithmic approaches and quantum circuit architectures, but they all share a general definition using common key elements [34]: a quantum circuit built on a set of variational parameters $\{\theta_i\}_i$, a cost operator $\hat{C}$, a loss function $\mathcal{L}$, and a classical optimizer.

#### 1. Variational quantum circuit

In the VQAs studied in this paper, the variational quantum circuits (VQCs) are composed of two parts: an encoding block, sometimes referred to as a quantum feature map circuit, and a variational *Ansatz*, sometimes directly referred to as the variational quantum circuit itself. While the encoding block is used to initialize the circuit with fixed parameters, the variational *Ansatz* is built on the set of variational parameters $\{\theta_i\}_i$ later optimized throughout the algorithm.

#### 2. Cost operator and loss function

One of the core features of a VQA is to select the architecture of the VQC and a suitable cost operator $\hat{C}$ for a given problem. In this paper, the cost operator is distinguished from the loss function $\mathcal{L}$. The cost operator maps observables to the VQCs, while the loss function guides the optimization process by quantifying the distance between the algorithm's outputs and the desired solution.

#### 3. Classical optimizers

Finally, the VQC is trained by a classical optimizer. As the loss function may have multiple local extrema, training a VQC is recognized as NP-hard [35] while facing new challenges due to the stochastic nature of quantum computation through measurements, noise, and presence of the barren plateau (detailed in Sec. II A 4) [34]. Consequently, the choice of a classical optimizer remains an active area of research.

One of the initial classical optimizers used for VQAs is the Adam optimizer, widely used in classical machine learning in the training of neural networks due to its ease of implementation, computational efficiency, and little memory requirements [36]. The Adam optimizer is an advanced version of the stochastic gradient descent (SGD) which uses first-order gradients to estimate first and second moments, thereby adapting the learning rate for each parameter iteratively throughout the optimization process. This optimizer finds application in the VQA by Kyriienko *et al.* [29], and is chosen for the VQA proposed in this paper, detailed in Sec. III. The second classical optimizer mentioned in this paper through the VQA proposed by Sato *et al.* [31] is the Broyden-Fletcher-Goldfarb-Shanno (BFGS) optimizer. While the Adam optimizer is designed to handle sparse gradients and noise, the BFGS aims to solve unconstrained nonlinear optimization problems. This optimizer is a quasi-Newton approach which approximates the Hessian matrix of the loss function from the first-order gradient or an approximation of the gradient.

The performance of optimizers for VQAs depends on the VQC structure and the loss function. A recent comparison between classical gradient-based and gradient-free optimizers [37] for the variational linear quantum solver [38] over 3 to 5 qubits with realistic noise did not highlight an optimizer type that outperformed the other. Among gradient-based optimizers, the best performances were obtained with the BFGS optimizer, while the overall best performances were obtained with the simultaneous perturbation stochastic approximation (SPSA) [39], a gradient-free optimizer. This optimizer is considered as gradient free as it approximates gradients from a single partial derivative obtained with finite differencing along a random direction. The SPSA is expected to be an efficient method for VQAs due to its reduced complexity and better performance compared to a SGD approach. However, other gradient-free optimizers such as COBYLA [40] performed poorly in the presence of noise compared to other optimizers [37].

Since these latter classical optimizers were initially proposed for classical computation, other methods have been developed recently for quantum hybrid computation, with, for example, the metalearner [41], a SGD-based optimizer, which adapts the learning rate of the optimization process by training a neural network based on the optimization history of similar problems. Another adaptive and SGD-based approach for VQA is to variate the number of shots required to determine the loss function gradient, that is, its precision, throughout the optimization process to significantly reduce the complexity of the optimization [42,43]. One more gradient-based approach for VQAs is the quantum natural gradient descent method, which instead evaluates the gradient of the loss function in terms of the Euclidean norms. It uses a metric tensor that quantifies the sensitivity of the quantum circuit to variation of the parameters [44]. Finally, in the case of a loss function defined as a sum of trigonometric functions, a gradient-free approach can be used where all the local parameters are sequentially updated [45,46].

#### 4. Barren plateau

The barren-plateau phenomenon refers to the problem of vanishing gradients in VQAs. It severely limits the trainability of these algorithms and poses a challenge for optimization problems in quantum systems. This phenomenon continues to receive a significant attention from researchers, to comprehensively understand the phenomenon, ascertain the reasons behind its occurrence, and explore effective strategies to surmount its challenges.

The first cause identified for the barren-plateau phenomenon is the use of deep randomly initialized hardware efficient *Ansätzes* [47]. Consequently, attention to overcome the barren plateau was initially directed toward a clever initialization of the VQC parameters [48–51] or a clever VQC structure [52]. Later, Cerezo *et al.* [32] studied the correlation between the depth of the circuit as well as the use of a local or global measurement and the occurrence of barren plateau. Their findings revealed that even shallow VQCs could lead to a barren plateau, highlighting the selection of the cost function as a main cause of the phenomenon. Specifically, opting for local cost functions over global cost functions was shown to mitigate the barren plateau, resulting in a polynomial gradient decay with circuit size instead of an exponential decay. Recently, one more approach proposed counteracting the emergence of the barren plateau by computing the gradient and variance of the cost function using Clifford approximant quantum circuits [33].

### B. Derivative quantum circuits

Analytical derivatives of quantum circuits with respect to Pauli gate parameters can be obtained easily by adding a corresponding rotational gate, depending on the circuit structure.

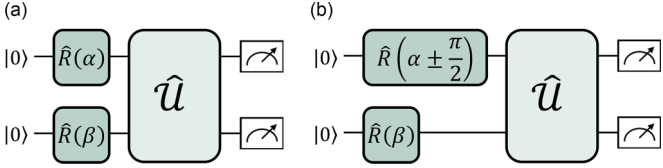

FIG. 1. (a) Original quantum circuit; (b) derivative quantum circuits.

#### 1. Parameter shift rule method

The parameter shift rule method is widely used to differentiate quantum circuits. This method requires to build two modified circuits to obtain the derivative with respect to one Pauli gate parameter [28]. In the example of a circuit of two qubits as described in Fig. 1(a), the derivative with respect to the parameter gate $\alpha$ is obtained by the difference of two modified versions where the parameter gate $\alpha$ is shifted to $\pm \pi/2$, as described in Fig. 1(b).

Thus, the derivative of the original quantum circuit in Fig. 1(a) with respect to $\alpha$ is

$$\frac{\partial}{\partial \alpha} \langle f_\alpha | \hat{C} | f_\alpha \rangle = \frac{1}{2}(\langle f_{\alpha^+} | \hat{C} | f_{\alpha^+} \rangle - \langle f_{\alpha^-} | \hat{C} | f_{\alpha^-} \rangle), \quad (1)$$

where $|f_\alpha\rangle$ is the state vector of the original quantum circuit, $|f_{\alpha^+}\rangle$ and $|f_{\alpha^-}\rangle$ are the state vectors of the quantum circuit where the parameter $\alpha$ has been shifted, respectively, of $\pi/2$ and $-\pi/2$. $\hat{C}$ is a chosen cost function.

#### 2. Hadamard test differentiation

A second method to derive quantum circuits is based on the Hadamard test, originally used to extract the real part of an expectation value by measuring the magnetization of the ancilla qubit $a$, as shown in Fig. 2(a): $\langle a | \hat{Z} | a \rangle = \text{Re}\langle \psi | \hat{\mathcal{U}} | \psi \rangle$. This structure can be reused to obtain the derivative of the expectation value real part with respect to a Pauli gate parameter $\theta_j$ [Fig. 2(b)] [53]:

$$\langle a | \hat{Z} | a \rangle = \frac{\partial}{\partial \theta_j} \text{Re}\langle \psi | \hat{\mathcal{U}} | \psi \rangle, \quad (2)$$

where $\hat{R}_j = \exp(\frac{-i}{2}\theta_j \hat{P}_j)$, and $\hat{P}$ is a tensor of Pauli matrices. With this differentiation method, only one circuit is required to obtain the derivative of the expectation value, but only the real value of the expectation is extracted.

### C. Kyriienko et al. algorithm

The first quantum algorithm studied in this work for comparison purposes is a VQA based on the algorithm developed by Kyriienko et al. [29], which trains VQCs to approximate the solution of DEs by polynomial fitting. In their paper,

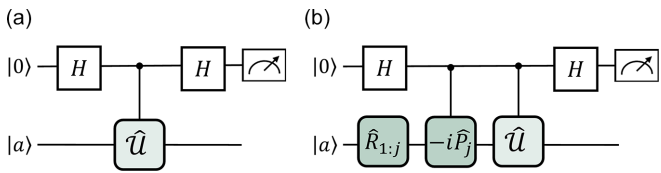

FIG. 2. (a) Hadamard test; (b) Hadamard test differentiation.

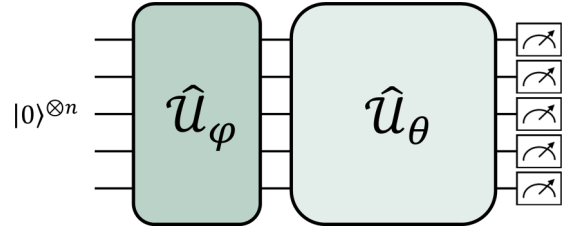

FIG. 3. Circuit structure of the Kyriienko et al. algorithm.

Kyriienko et al. applied their algorithm exclusively to first-order DEs. However, they suggested the possibility of extending its use to higher-order DEs through repeated circuit differentiation. In this work, we propose an extension of the algorithm's application to second-order DEs.

#### 1. Circuit structure

The Kyriienko et al. algorithm [29] aims to represent an approximation of the solution of a differential equation by a parametrized quantum circuit divided into two parts: a quantum feature map $\hat{\mathcal{U}}_\varphi(x)$ and a variational *Ansatz* $\hat{\mathcal{U}}_\theta$, as shown in Fig. 3. The first part encodes the equation variable $x$ through a fitting set of nonlinear Chebyshev polynomials of the first and second kind over ]0,1[. Encoding the variable $x$ through functions allows a high expressivity of the quantum circuit while ensuring its derivability. The quantum feature map $\hat{\mathcal{U}}_\varphi(x)$ is composed by $Y$-rotational gates $\hat{R}_Y$, parametrized by the functions $\varphi_j(x) = 2j \arccos(x)$, where the degree $j$ of the Chebyshev polynomials grows with the number of qubits $n$ [29]. $\hat{\mathcal{U}}_\varphi(x)$ can be expressed as

$$\hat{\mathcal{U}}_\varphi(x) = \bigotimes_{j=1}^{n} \hat{R}_{Y,j}[2j \arccos(x)]. \quad (3)$$

This quantum feature map encodes the Chebyshev polynomials of the first and second kind as follows:

$$\hat{R}_{Y,n}(\varphi(x))|0\rangle = T_n(x)|0\rangle + \sqrt{1-x^2}\, U_{n-1}(x)|1\rangle, \quad (4)$$

where the Chebyshev polynomials of the first kind are defined by $T_n = \cos[n \arccos(x)]$ and of the second kind by $U_{n-1}(x) = \sin[n \arccos(x)]/\sin[\arccos(x)]$ [29]. The chosen variational *Ansatz* $\hat{\mathcal{U}}_\theta$ is the hardware efficient *Ansatz* [54], composed of multiple layers of a sequence of rotational gates and controlled-NOT (CNOT) gates.

#### 2. Algorithm

In Kyriienko et al. [29], the resulting quantum state $|f_{\varphi,\theta}(x)\rangle = \hat{\mathcal{U}}_\theta \hat{\mathcal{U}}_\varphi(x)|\emptyset\rangle$ is evaluated by the total magnetization of the system $\hat{C} = \sum_{j=1}^{n} \hat{Z}_j$. The derivatives with respect to $x$ and $\theta$ are determined by the parameter shift rule method and the chosen classical optimizer is Adam [36]. Finally, the loss function is defined as the mean square of the DE with all terms on one side, as expressed in Eq. (6).

### D. Sato et al. algorithm

The second VQA used in this work for comparison purposes is the Sato et al. algorithm [31], which was designed to solve the Poisson equation. This algorithm is based on a discretized approach in which the VQC is used to approximate the unitary vector of the solution of the 1D Poisson equation.

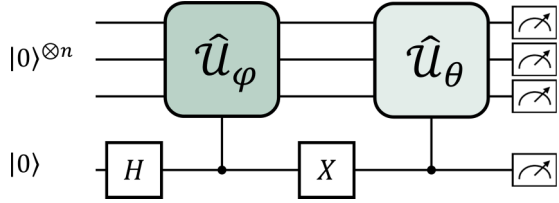

FIG. 4. Circuit structure of the Sato *et al.* algorithm.

### 1. Circuit structure

The circuit structure has been built to solve a 1D second-order differential equation of the form $d^2 f(x)/dx^2 + s(x) = 0$, where $s$ is the source term encoded through its amplitudes by $\hat{\mathcal{U}}_s$ and the unitary vector of the solution $f$ is approximated by the variational *Ansatz* $\hat{\mathcal{U}}_\theta$. This structure using an ancilla gives the following state:

$$|s, \psi_\theta\rangle = \frac{1}{\sqrt{2}}(|0\rangle|s\rangle + |1\rangle|\psi_\theta\rangle), \quad (5)$$

where the state vector $|\psi_\theta\rangle$ encodes the unitary vector of the discretized vector solution $f$.

### 2. Algorithm

This approach evaluates the VQC described in Fig. 4 with multiple cost operators depending on the boundary conditions to quantify the loss function, which here is the potential energy of the DE. As this algorithm is a discretized approach, derivatives with respect to the variable $x$ are obtained with a finite-difference scheme. In contrast, the gradient of the loss function $\nabla_\theta \mathcal{L}_\theta$ requires differentiating the VQC, using a method based on the Hadamard test differentiation [31]. The classical optimizer chosen by Sato *et al.* is the BFGS optimizer, described in Sec. II A 3.

## III. METHODS

### A. Algorithms workflow

For both algorithms studied here, the quantum circuits are aimed to approximate the solution of a DE. To do so, the algorithms follow a workflow inspired by Kyriienko *et al.* [29], summarized in Fig. 5. First, an initial set of quantum circuits and their derivatives with respect to $x$ is built for each point $x_i$ in the chosen interval of interest $I$, with a random vector of variational parameters $\{\theta_i\}$: this is the initial state of the VQA. The value of the resulting function is read as an expectation value through a cost operator $\hat{C}$ applied to the quantum state $|f_{\varphi,\theta}(x)\rangle$ of the circuit: $f_\theta(x) = \langle f_{\varphi,\theta}(x)|\hat{C}|f_{\varphi,\theta}(x)\rangle$. The classical optimizer uses a loss function, which evaluates if the resulting function satisfies the DE as well as other conditions like the initial or boundary conditions of the given problem or regularization points. In this work, the selected classical optimizer is Adam [36], which demonstrated satisfactory performances in previous studies [29]. As this optimizer relies on gradient descent optimization, derivative quantum circuits with respect to the variational parameters $\theta$ must be calculated. Finally, the vector of variational parameters $\theta$ is updated and all circuits are recalculated in another loop until the resulting function converges to the desired function.

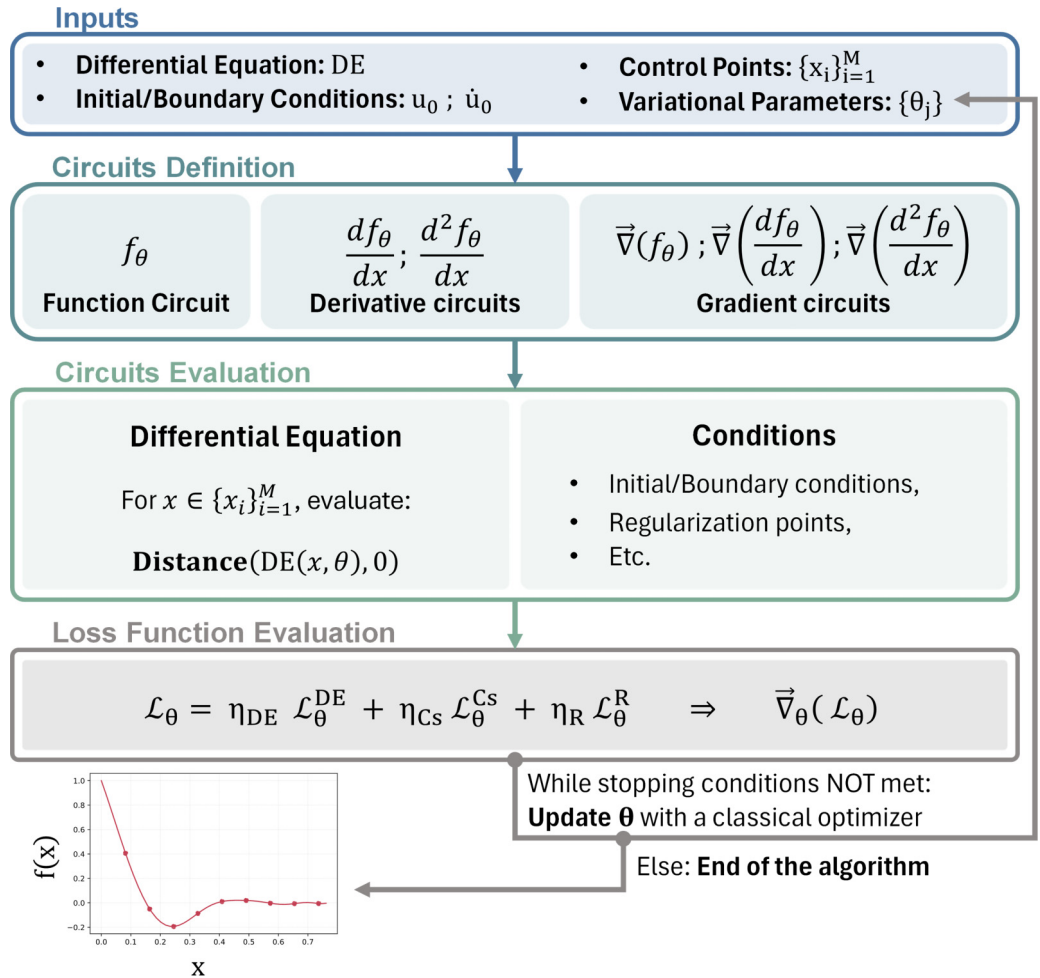

FIG. 5. Proposed algorithm workflow.

This work specifically focuses on enhancing the VQC structure to not only encode other polynomials but also to effectively determine the gradient and derivatives at a lower computation cost by comparison to existing approach while not compromising on the quality of the solution, thereby making this new algorithm attractive.

### B. Loss function

With a similar approach to the Kyriienko *et al.* algorithm [29], the loss function $\mathcal{L}_\theta^{\mathrm{DE}}$ quantifies the distance between the DE, written with all terms on one side, and zero, over a set of $M$ points within the chosen interval $I$. This loss function is defined as

$$\mathcal{L}_\theta^{\mathrm{DE}}\left[\frac{\partial^2 f_\theta}{\partial x^2}, \frac{\partial f_\theta}{\partial x}, f_\theta, x\right]$$
$$= \frac{1}{M}\sum_{i=1}^{M} L\left(\mathrm{DE}\left[\frac{\partial^2 f_\theta}{\partial x^2}(x_i), \frac{\partial f_\theta}{\partial x}(x_i), f_\theta(x_i), x_i\right], 0\right), \quad (6)$$

where the distance $L$ is evaluated with the mean-square error $L(a,b) = (a-b)^2$. Other distance evaluations can be used, such as the mean absolute error $L(a,b) = |a-b|$. This loss function can be completed with additional functions that quantify the satisfaction of initial or boundary conditions $\mathcal{L}_\theta^{Cs}$ or regularization points $\mathcal{L}_\theta^R$ [29]. Arbitrary weights $\eta$ can also be used to prioritize a condition over the others, as expressed in the equation

$$\mathcal{L}_\theta = \eta_{\mathrm{DE}}\,\mathcal{L}_\theta^{\mathrm{DE}} + \eta_{Cs}\,\mathcal{L}_\theta^{Cs} + \eta_R\,\mathcal{L}_\theta^R. \quad (7)$$

### C. Initial and boundary conditions

In this work, the differential equation can be either an initial value problem (IVP) or a boundary value problem (BVP). In both cases, the associated *initial* or *boundary conditions* represent constraints that the solution must satisfy. These conditions can be taken into account in the optimization loop by contributing to the loss function, or can also be handled separately as an automatic shift of the function read out from the quantum circuit as proposed by Kyriienko *et al.* [29].

#### 1. Contribution to the loss function

For a second-order DE, the contribution of the initial or boundary conditions to the loss function can be described as follows:

$$\mathcal{L}_\theta^{Cs} = L(f_\theta(x_0), u_0) + L\left(\frac{\partial f_\theta}{\partial x}(x_0), \dot{u}_0\right), \quad (8)$$

where $u_0$ and $\dot{u}_0$ are the value of the sought function and its derivative at $x = x_0$. For a lower- or higher-order DE, the corresponding terms must be removed or added.

#### 2. Floating boundary handling

As an alternative approach, a shift term $f_{Cs}$ can be calculated at each iteration to automatically match the initial or boundary condition regardless of the value of the variable parameters $\theta$ [29]. Hence, the contribution to the loss function is null ($\mathcal{L}_\theta^{Cs} = 0$), and we have

$$f_\theta(x) = \langle f_{\varphi,\theta}(x)|\hat{C}|f_{\varphi,\theta}(x)\rangle + f_{Cs}, \quad (9)$$

$$f_{Cs} = u_0 - \langle f_{\varphi,\theta}(x_0)|\hat{C}|f_{\varphi,\theta}(x_0)\rangle. \quad (10)$$

This definition can be extended for the derivatives of the sought function for all DEs. However, to ensure the flexibility of the algorithm and its convergence toward the sought function, the initial or boundary condition can be handled with a shift term for the function and a loss contribution for the derivatives.

### D. Regularization

For equations where some values of the desired function $\{u_r\}$ are known, an additional condition can be created to ensure the convergence of the desired function toward those values. This contribution to the loss function can be defined as

$$\mathcal{L}_\theta^R = \frac{1}{M_R}\sum_{r=1}^{M_R}(L(f_\theta(x_r), u_r)), \quad (11)$$

where $M_R$ is the number of regularization points and $u_r$ the value of the sought function at those points. A regularization procedure can also be implemented for derivatives values [29]. To minimize the complexity of the algorithm, regularization can be added to the loop by dividing the set of points $\{x_i\}_{i=1}^{M}$ into subsets of points that will, one by one, be used as training points for the differential equation and then as regularization points once the algorithm converged on this subset. This approach can save an important amount of computing resources but can in certain cases decrease the overall accuracy of the algorithm output. In the first application presented in Sec. IV A, regularization has been used for the Hadamard-Lagrange algorithm, to reduce its overall gate complexity.

### E. Cost operator

The cost operator $\hat{C}$ is used to read out the value of the approximated function and its derivatives from the different quantum circuits built throughout the algorithm, with

$$f(x) = \langle f_{\varphi,\theta}(x)|\hat{C}|f_{\varphi,\theta}(x)\rangle, \quad (12)$$

where $|f_{\varphi,\theta}(x)\rangle$ is the quantum state of the quantum circuit evaluated at $x$ within the chosen interval. The choice of the cost operator impacts the flexibility and trainability of the algorithm. Most common cost functions are the magnetization of a single qubit $j$: $\hat{C} = \hat{Z}_j$, which can be used for the measurement of an ancilla qubit, for example, or the total magnetization of the system, where all qubits are measured: $\hat{C} = \sum_{j=1}^{n} \hat{Z}_j$.

In this study, the cost functions employed are centered around the total magnetization of the system, or variations of it. The decision to opt for projective measurement along the $Z$ axis is motivated by its straightforward implementation and the potential to simplify the variational *Ansatz*, reducing the overall gate complexity of VQAs as detailed in Appendix A.

### F. Derivatives and gradients

The two new VQA architectures proposed in this paper aim to represent the function solution of the DE through a quantum circuit. To do so, the derivatives with respect to the variable $x$ are necessary to evaluate the loss function. Both quantum feature maps encode the variable $x$ using a set of functions $\{\varphi_i\}_{i=0}^{n}$ where $n$ is the number of gates used for the quantum feature map encoding. Hence, the first- and second-order derivatives are defined as follows:

$$\frac{df}{dx} = \sum_{i=0}^{n} \frac{\partial}{\partial \varphi_i} \langle f_{\varphi,\theta,i} | \hat{C} | f_{\varphi,\theta,i} \rangle \frac{d\varphi_i}{dx_i}, \quad (13)$$

$$\frac{d^2 f}{dx^2} = \sum_{i=0}^{n} \frac{\partial}{\partial \varphi_i} \langle f_{\varphi,\theta,i} | \hat{C} | f_{\varphi,\theta,i} \rangle \frac{d^2 \varphi_i}{dx_i^2}$$
$$+ \sum_{i=0}^{n} \sum_{j=0}^{n} \frac{\partial^2}{\partial \varphi_i \varphi_j} \langle f_{\varphi,\theta,i,j} | \hat{C} | f_{\varphi,\theta,i,j} \rangle \frac{d\varphi_i}{dx_i} \frac{d\varphi_j}{dx_j}. \quad (14)$$

However, in the variational *Ansatz*, the variational parameter $\theta$ is encoded directly into the parametrized rotational gates. In this work, all gradients $\nabla_\theta$ are determined using the parameter shift rule method defined in Eq. (1).

### G. Iterative process

The iterative process used in this paper to train the VQAs uses the *Adam* classical optimizer [36]. Adam is a gradient-based optimization algorithm that combines the advantages of momentum and adaptive learning rates for efficient parameter updates. It iteratively minimizes the loss function by adjusting the trainable parameters $\theta$, based on estimates of first- and second-order moments of the gradients. To ensure a robust and efficient optimization process, this work employed the convergence criteria on the loss function and its gradient. The iterative process halts if the loss function reaches a predefined target corresponding to the desired accuracy $\epsilon_\mathcal{L}$. Alternatively, the algorithm is considered to have converged when the absolute value of each component of the loss function gradient falls below a specified threshold of $\epsilon_{\partial \mathcal{L}}$:

$$\left| \frac{\partial \mathcal{L}}{\partial \theta_i} \right| \leqslant \epsilon_{\partial \mathcal{L}}. \quad (15)$$

The gradient criterion prevents unnecessary computations once the solution is stabilized, while the loss function value criterion guarantees accuracy. Together, these criteria effectively balance efficiency and performance during optimization.

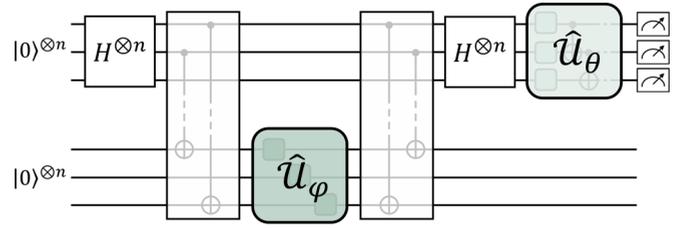

FIG. 6. Circuit structure of the Hadamard-Lagrange algorithm with an extended structure.

### H. Quantum circuit structures

For the two new VQA architectures, the variable function $x$ is encoded through a quantum feature map circuit $\hat{\mathcal{U}}_\varphi(x)$ and the vector of variational parameters $\theta$ is encoded through a variable *Ansatz* $\hat{\mathcal{U}}_\theta$. The VQA inspired by the work of Kyriienko *et al.* [29] is based on two circuits acting on the same register, one after the other, and all qubits in the register are measured, as shown in Fig. 3. For the newly designed VQA, two different structures have been studied (Figs. 6 and 7) and in both cases the circuit is inspired by the Hadamard test structure (Fig. 2), over the ancilla register of one qubit or multiple qubits.

The development of the Hadamard-Lagrange algorithm has been driven by two primary factors. First, due to the computational demands of the parameter shift rule method, the gate complexity involved in determining the first and second derivatives with respect to $x$, along with their gradients with respect of $\theta$, was substantial. Second, observations from the initial application of the Kyriienko-inspired algorithm to a second-order DE, as outlined in Sec. IV A, revealed oscillations which may be caused by the use of excessively high-degree polynomials for interpolation tasks. Consequently, this study focused on a VQC structure that not only reduces the computational overhead associated with evaluating derivatives and gradients, but also implementing the Lagrange interpolation, while not impacting the quality and accuracy of the solution. One of the main objectives was also to perform the interpolations with the lowest possible degree of polynomial.

The first approach was initially designed across two quantum registers of the same sizes to ensure the VQC's trainability. As discussed in Sec. II A 4, local measurements can mitigate the barren-plateau phenomenon. However, for circuits with very shallow depths, this extended structure is suboptimal as it necessitates a high number of qubits. Therefore, the Hadamard-Lagrange structure was redesigned to minimize the required number of qubits for encoding, while retaining the same cost function and differentiation method (second approach). Intermediate structures where the number of qubits required for the encoding is between 1 and $n$ could also have been considered, but the study of such structure is left for a future work.

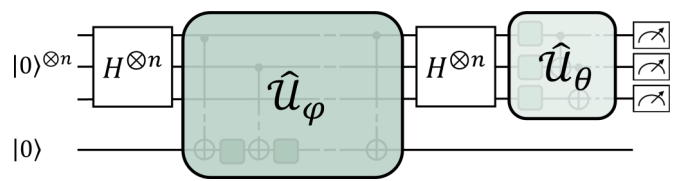

FIG. 7. Circuit structure of the Hadamard-Lagrange algorithm with a simplified structure.

### I. Kyriienko-inspired algorithm for comparison

In order to assess the performance of our new VQA, the algorithm developed by Kyriienko *et al.* [29] is expanded following Eq. (14). The hardware-efficient *Ansatz* originally used for the variational *Ansatz* is replaced with two layers of single-qubit $X$-rotational gates $\hat{R}_X$ per qubit and a linear network of CNOT gates. This modification reduces the complexity of the classical optimizer since the $Z$-rotational gates $\hat{R}_Z$ alone do not affect projective measurements along the $Z$ axis (see Appendix A for more details).

### J. Proposed new Hadamard-Lagrange algorithm

The new algorithm designed and tested in this work has been developed based on the Hadamard test structure (Fig. 2) but extended to a multiple-qubit register, to approximate the solution of the differential equation by Lagrange interpolation. In this work, two different structures of the quantum feature map have been proposed.

#### 1. Quantum feature map

First, an extended version has been developed where an encoding block $\hat{\mathcal{U}}_\varphi$, with $\varphi_i(x) = \arccos(\frac{x-x_j}{2})$, is applied in-between two networks of CNOT gates, over a quantum register of $n$ qubits, leading to

$$\hat{\mathcal{U}}_\varphi(x) = \bigotimes_{j=1}^{n} \hat{R}_{Y,j}\left[\arccos\left(\frac{x-x_j}{2}\right)\right]. \quad (16)$$

A second version with a simplified structure has been developed, where only one ancilla qubit is used for the encoding. Here, the variable $x$ is encoded through $Y$-rotational gates parametrized by the same encoding function $\varphi_i$ defined in the extended structure but within the networks of CNOT gates:

$$\hat{\mathcal{U}}_\varphi(x) = \left(\prod_{i=1}^{n-1} \text{CNOT}_{i,a}\, \hat{R}_{Y,a}[\varphi_{i+1}(x)]\right)$$
$$\times \text{CNOT}_{n,a}\left(\prod_{i=1}^{n-1} \text{CNOT}_{i,a}\, \hat{R}_{Y,a}[\varphi_i(x)]\right)\text{CNOT}_{n,a}, \quad (17)$$

where $\text{CNOT}_{i,a}$ denotes a controlled-NOT gate with control on the $i$th qubit and target on the ancilla qubit, and $\hat{R}_{Y,a}(\varphi)$ represents a rotation around the $Y$ axis applied to the ancilla qubit, defined as $\hat{R}_{Y,a} = I^{\otimes n} \otimes R_Y$.

In both structures, the set of points $\{x_j\}_{j=1}^n$ included in the interval $I$ used for the polynomial interpolation is directly linked to the size of the first quantum register. While the second register can be built over $n$ qubits (Fig. 6) or only one qubit (Fig. 7), for both structures, only the first register is measured where the observable corresponds to the magnetization of the first $n$ qubits, expressed as a sum of $Z_i$ operators, leading to local measurements instead of global measurements. As discussed in Sec. II A 4, this allows to delay the appearance of the barren-plateau phenomenon.

Additionally, determining whether to use a simplified or extended circuit structure depends on the quantum device being employed. The extended circuit structure requires more qubits but fewer gates in comparison to its simplified counterpart (described in Appendix C). Specifically, the simplified circuit structure saves $n - 1$ qubits, but requires $\lfloor n/2 \rfloor$ additional gates. Consequently, while it reduces noise from gate errors or decoherence, the extended structure may be prone to errors from qubits or crosstalk, and can possibly be limited by the existing quantum resources available in NISQ devices. Conversely, the simplified structure shows promise, particularly with the development of efficient mitigation techniques and error gate correction. In the realm of quantum circuit simulations, the preference leans towards the simplified circuit structure due to its lower computational resource demands (as both circuit provides the same solution on noise-free qubits).

The choice of the encoding function $\varphi_i = \arccos(\frac{x-x_i}{2})$, and specifically the introduction of the factor $\frac{1}{2}$, is motivated by the behavior of its derivatives. While the difference $(x - x_i)$ naturally varies within the interval $]-1, 1[$, which matches the domain of definition for the arccos function, both derivatives diverge when $|(x - x_i)|$ approaches 1:

$$\frac{d}{d\gamma}\arccos(\gamma) = -\frac{1}{\sqrt{1-\gamma^2}},$$
$$\frac{d^2}{d\gamma^2}\arccos(\gamma) = -\frac{\gamma}{(1-\gamma^2)^{3/2}}. \quad (18)$$

If the argument were chosen as $(x - x_i)$ directly, this divergence would occur more frequently within the range of interest, potentially leading to numerical instabilities during optimization. Introducing the $\frac{1}{2}$ scaling ensures that $|\gamma| = |\frac{x-x_i}{2}|$ remains sufficiently below 1, thus avoiding these divergences and improving the numerical stability of the encoding.

#### 2. Variational quantum circuit

In our new VQA, the variational *Ansatz* $\hat{\mathcal{U}}_\theta$ is composed of one layer, following the same structure as presented in the Kyriienko-inspired algorithm which is used for comparison (see Sec. III I).

#### 3. Cost operator

For both circuit structures to be read as Lagrange interpolating polynomial, the first register is evaluated by the same cost operator:

$$\hat{C} = \sum_{j=1}^{n} \frac{1}{\rho_j} \hat{Z}_j, \text{ with } \rho_j = \langle f_{\varphi,0}(x_j)|\hat{Z}_j|f_{\varphi,0}(x_j)\rangle. \quad (19)$$

The resulting function is a Lagrange interpolating polynomial:

$$f(x) = \sum_{j=1}^{n} \left(\alpha_\theta \prod_{i \neq j}^{n} \frac{x - x_i}{x_j - x_i}\right), \quad (20)$$

where $\alpha_\theta$ is a coefficient determined by the variational *Ansatz*. The detail of the calculation is given in Appendix B.

## IV. RESULTS

In order to assess the performance of our new Hadamard-Lagrange algorithm, two DEs will be used, as well as existing VQAs and analytical solutions for comparisons, which will be performed with a focus on two metrics: the quality of the solution and the gate complexity of the VQAs.

### A. Damped mass-spring system

A damped mass-spring system (DMSS) is a physical model used to describe the behavior of a mass attached to a spring with damping. This system is widely studied in mechanics and physics to understand oscillatory motion and damping effects. In this work, we focus on an initial value problem (IVP) of a DMSS solved over a time interval of $I = [0, 10]$ s, using the Kyriienko-inspired algorithm and the proposed new Hadamard-Lagrange algorithm with a reduced circuit structure. The DE can be expressed as

$$m\frac{d^2 f}{dt^2} + b\frac{df}{dt} + kf = 0, \quad (21)$$

with $m = 1$ kg; $b = 1$ kg s$^{-1}$; $k = 1$ N m$^{-1}$ and the initial conditions of the DE are $u_0 = 1$ m and $\dot{u}_0 = 0$ m s$^{-1}$, which will be treated similarly to boundary conditions (BC) as explained in Sec. III C. The results obtained by the two quantum methods are compared with the analytical solution, detailed in Appendix D.

In this study, two quantum approaches are applied to the damped mass-spring system differential equation and analyzed by looking at their performance using two distinct sets of training points. These sets are based on Chebyshev nodes, recognized as optimal for avoiding the Runge phenomenon [55] when interpolating over equispaced points induces oscillations at the edges of the interval [55]. The first set $\{x_i^{(1)}\}_i$ corresponds to the original Chebyshev nodes (originally between $[-1, 1]$) rescaled over an interval $[a, b]$ (with $0 \leqslant a < b \leqslant 1$). The second set $\{x_i^{(2)}\}_i$ corresponds only to the original positive Chebyshev nodes scaled over an interval $[0, b]$. Here, $a = 0$ and $b = 0.9$ as imposed by qubits encoding and for stability reason, for both sets of points. These two sets can be expressed as

$$x_k^{(1)} = \frac{a+b}{2} + \frac{b-a}{2}\cos\left(\frac{2k-1}{2N_{\text{points}}}\pi\right), \quad 0 \leqslant k < N_{\text{points}} \quad (22)$$

$$x_k^{(2)} = b\cos\left(\frac{2k-1}{4(N_{\text{points}}-1)}\pi\right), \quad 0 \leqslant k < N_{\text{points}}. \quad (23)$$

To assess the quality of the result and facilitate a comparative analysis of the two quantum algorithms, we will differentiate the DE loss contribution at each individual point, represented by the square of the DE evaluated at that point, and, the total loss as the average of DE loss contributions across all points considered, as defined in Eq. (7). Here, the loss is assessed over 50 equispaced points within the specified interval.

To estimate the gate complexity between the two quantum approaches, the number of circuits and basic quantum gates required for one iteration and the entire calculation are compared. The detailed methodology for estimating this gate complexity is provided in Appendix C. However, these results can be highly dependent on the initial random values of the vector $\theta$, providing only an estimate of the order of magnitude of the quantum circuits and quantum gates required to solve the equation.

#### 1. Kyriienko-inspired algorithm

The application of the Kyriienko-inspired (KI) algorithm to the DMSS problem has been investigated using a range of VQCs with 3 to 5 qubits and 2 to 7 layers, across 7, 9, or 12 points for both sets defined in Eqs. (22) and (23). Each VQC is initialized with a random vector of variational parameters $\theta$. To ensure a fair comparison of the influence of the size and type of training set, the same initial vector $\theta$ has been used for a given size and depth of the VQC. Furthermore, the quantum feature map used in the KI algorithm is based on high-degree Chebyshev polynomials, which can cause oscillations at the extremities of the algorithm. To mitigate this issue, the KI algorithm has been trained over an interval of [0,12] s, ensuring a smooth solution over the required interval of [0,10] s.

Additionally, the quantum feature map of this algorithm's structure is independent of the training set of points, and increasing the number of training points throughout the algorithm does not guarantee a more efficient optimization. In fact, more iterations may be necessary to achieve convergence. Hence, the VQCs have been trained over all points in the chosen set, using the Adam classical optimizer [36] with a learning rate of $\alpha = 0.01$.

Convergence was determined using the criteria outlined in the Methods section: a gradient threshold of $\epsilon_{\partial \mathcal{L}} = 10^{-4}$, as defined in Eq. (15), or achieving the desired loss function value corresponding to the target accuracy $\epsilon_{\mathcal{L}} = 10^{-4}$. The choice of the threshold value $\epsilon_{\partial \mathcal{L}} = 10^{-4}$ is based on prior experience and empirical observations, where it was found that no significant changes occur in the optimization process once all gradient components are smaller than this value.

For this application, the initial condition is implemented with a floating boundary handling (Sec. III C 2) for $u_0$, in combination with a contribution to the loss function (Sec. III C 1) for $\dot{u}_0$. This ensures the satisfaction of the first condition while avoiding to overconstrain the system. In this case, the loss function can be expressed by Eq. (7), with $\eta_{\text{DE}} = 1$, $\eta_{Cs} = 1$, and $\eta_R = 0$.

Among the different VQCs studied for the KI algorithm, the lowest average DE losses were observed with a VQC of 5 qubits and 2 layers over a set of 12 points of the second kind (referred as *N5L2I2*) and with a VQC of 4 qubits and 3 layers over a set of 12 points of the first kind (referred as *N4L3I1*). According to the results obtained from this algorithm, an increased number of points in the training set does not always correlate with a higher gate complexity or highly improved accuracy. However, the selection of the training points themselves and the structure of the VQC seems to have the most significant impact on both the quality of the solution and the gate complexity, as shown in Fig. 8.

Additionally, applying this algorithm with different random initial vectors of variational parameters $\theta$ revealed a high sensitivity to these initial conditions. Across five independent attempts to solve the DMSS problem using the structures presented above, the $N5L2I2$ configuration exhibited a coefficient of variation of 20% in total loss and 57% in the number of iterations required for convergence. Similarly, for the $N4L3I1$ structure, the coefficient of variation was 50% in total loss and 40% in iterations. Consequently, the accuracy and gate complexity shown in Fig. 8 represent average outcomes, rather than the best or worst results achieved.

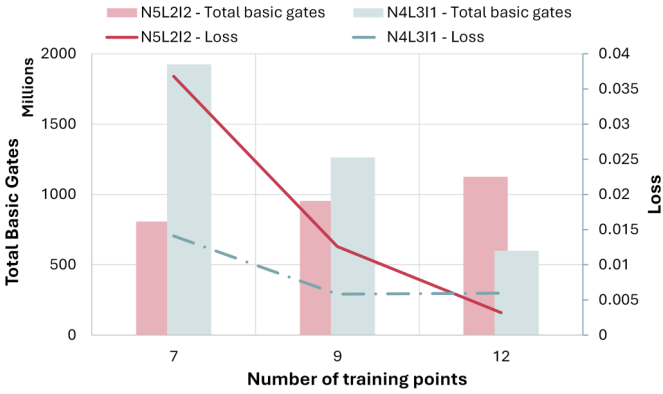

FIG. 8. Average total loss and gate complexity for the Kyriienko-inspired algorithm applied to the most efficient VQC structures with varying training set lengths over 5 tries. $N5L2I2$ stands for the structure of 5 qubits and 2 layers across a training set of the second kind, while $N4L3I1$ stands for the structure of 4 qubits and 3 layers across a training set of the first kind.

The variability in accuracy is further visualized in Fig. 9, where the best solution achieved for each VQC structure is plotted alongside a shaded region. This shaded region allows the visualization of the worst-case outputs for each structure, effectively capturing the range of variability across the attempts.

For the two VQC structures that produced the best results, some results presented important oscillations toward the end of the interval, despite placing some training points outside the interval to mitigate this issue. Additionally, for both structures, the initial (or boundary) conditions for the first- and second-order derivatives were not satisfied. Finally, the corresponding total DE loss obtained for the best results highlighted in Fig. 9 is, respectively, $2.50 \times 10^{-3}$ for $N5L2I2$ and $2.77 \times 10^{-3}$ for $N4L3I1$. However, the initial conditions for the first- and second-order derivatives are not fully met, resulting in a BC loss of 0.447 and 0.485 for each VQC structure, respectively, as illustrated in Fig. 10.

### 2. Hadamard-Lagrange algorithm

The new algorithm proposed in this work introduces the use of interpolation points in the encoding part of the VQC. These interpolation points are selected to be similar to the training set of points. This approach avoids overconstraining the optimization process and helps reduce the overall gate complexity. This new method has been studied using sets of 5 to 8 points. Among these, the optimal VQC size was found to be based on 7 points, corresponding to the use of 8 qubits, with the use of the simplified structure as illustrated in Fig. 7. As already mentioned, the focus of this work is on the theoretical aspects of variational quantum algorithms, hence, why the quantum circuits examined in this work are simulated in a noise-free environment. Consequently, the use of the simplified structure is preferred because of its lower computational resource demands.

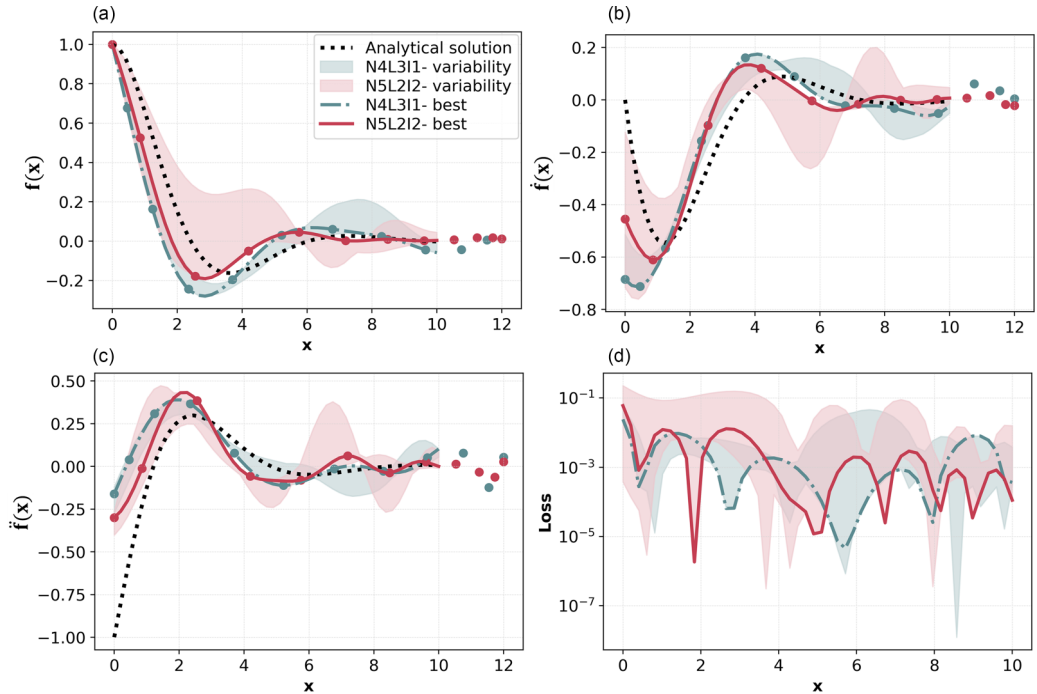

FIG. 9. Results are presented for each of the most efficient VQC structures, illustrating variability and the best outcome achieved across five attempts for (a) the function $f$, (b) the first derivative $\dot{f}$, (c) the second derivative $\ddot{f}$, and (d) the DE loss contribution across 50 equispaced points in the interval, plotted on a logarithmic scale. $N5L2I2$ stands for the structure of 5 qubits and 2 layers across a training set of the second kind, while $N4L3I1$ stands for the structure of 4 qubits and 3 layers across a training set of the first kind.

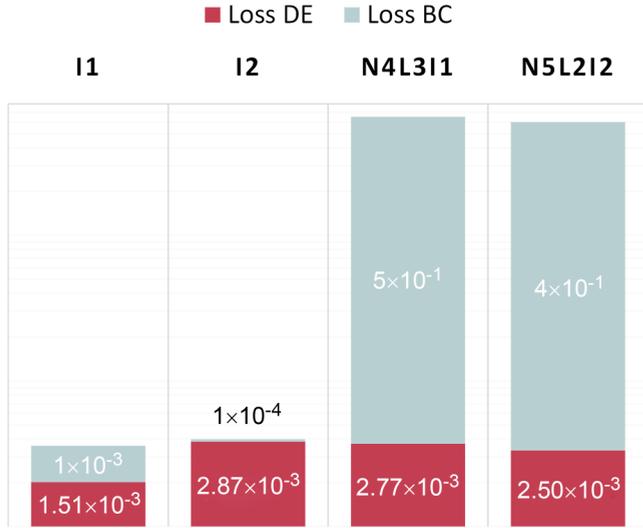

FIG. 10. DE and BC loss comparison between the new Hadamard-Lagrange VQA and the best outcomes of the Kyriienko-inspired VQA for both kinds of training sets of points, plotted on a logarithmic scale.

It should be noted that, for this algorithm, training points in pairs with the number of qubits can be modified during the calculation. To accelerate the optimization, the process is divided into two parts, using an evolving set of training points and regularization points. Furthermore, the initial (or boundary) conditions are implemented following the same methodology as for the Kyriienko-inspired algorithm, with this time $\eta_{\text{DE}} = 1$, $\eta_{Cs} = 0.6$, and $\eta_R = 1$.

For the first part of the optimization, the set of training points initially contains only the three Chebyshev nodes on the left of the domain (using four qubits). The first node contributes to the BC loss function $\mathcal{L}_\theta^{\text{BC}}$ and the other two points contribute to the DE loss function $\mathcal{L}_\theta^{\text{DE}}$. Once the algorithm converged for these two nodes, which means that all components of the loss function gradient reached a threshold value of $10^{-4}$ [defined in Eq. (15)] the next node is added and will contribute to the DE loss function $\mathcal{L}_\theta^{\text{DE}}$, while the second point will become a regularization point contributing to the regularization loss function $\mathcal{L}_\theta^R$. This process is repeated until convergence is achieved for all nodes (using eight qubits in this example), using the Adam classical optimizer [36] with an evolving learning rate from $\alpha = 0.04$ to $0.01$ depending on the total loss function value $\mathcal{L}_\theta$.

In the second part of the optimization, the size of the VQC does not evolve and the nodes contribute three by three to the DE loss, while other points act as regularization points, starting again from the left side of the interval. During this second part, the Adam classical optimizer [36] is applied with a fixed learning rate of $\alpha = 0.01$.

The application of this new VQA, using this two-part optimization, has been observed to be insensitive to the initial random vector of variational parameters $\theta$. For each type of training set, the solutions obtained after the first part of the optimization over five trials with different initial random values of the vector of variational parameters $\theta$, were extremely consistent, with a mean total loss of $1.24 \times 10^{-2}$ and $2.76 \times 10^{-2}$ with a coefficient of variation of 1.0% and 1.5% for the first and second kinds of training set, respectively. Similarly, the average number of iterations to achieve the convergence of the first part is, respectively, 670 and 620 iterations with a coefficient of variation of 11% in both cases. The second part of the optimization process consistently starts from a similar state, resulting in comparable deviation in terms the solution accuracy and the number of iterations.

As illustrated in Fig. 11, the first part of the optimization yielded a close approximation of the solution, achieving slightly better accuracy for the initial set of training points. Similarly, the second part of the optimization process produced improved results for the first training set, with a total DE loss of, respectively, $1.51 \times 10^{-3}$ and $2.87 \times 10^{-3}$. Additionally, the initial (or boundary) conditions are more effectively met this time, with a BC loss of, respectively, $1.18 \times 10^{-3}$ and $1.07 \times 10^{-4}$, as depicted in Fig. 10.

### 3. Hadamard-Lagrange algorithm vs Kyriienko-inspired algorithm

For both quantum algorithms, the number of qubits and Chebyshev nodes is different. The idea here was to select, by trial and error, the optimal set of training points and structures to obtain the most accurate solution for each approach.

As shown in Fig. 12, both quantum algorithms provide a satisfactory approximation of the desired function $f_{\theta_{\text{opt}}}$ over the chosen interval, with better accuracy with the Hadamard-Lagrange algorithm. Noticeable oscillations in the general approximation of the derivatives with respect to $x$ can be seen for the Kyriienko-inspired algorithm, despite training being performed over Chebyshev nodes (different locations for the nodes have been tested to confirm that the oscillations were still present). Additionally, for this algorithm, the initial conditions are not met for the derivatives, suggesting that the contribution of the loss function for the initial values may conflict with the overall loss function. It suggests that this quantum algorithm may lack flexibility to take into account all types of initial (or boundary) conditions, potentially due to constraints in the encoding or variational *Ansatz*.

The Hadamard-Lagrange algorithm reproduces the Lagrange interpolation and allows the determination of the fitting polynomial with the smallest degree over the given set of points. Hence, the function as well as their derivatives are smooth and do not present much oscillations in comparison with the Kyriienko-inspired algorithm. Moreover, the algorithm better adheres to the boundary conditions. In general, the solutions obtained via our new Hadamard-Lagrange algorithm appear to be of better quality compared to the solutions obtained with the Kyriienko-inspired algorithm, especially at the boundaries, and are in reasonably good agreement with the analytical solution as seen in Fig. 12. Furthermore, regarding the different loss contribution compared in Fig. 10, both algorithms converge toward a solution with a similar range of DE loss, but the contribution of the boundary conditions to the loss function considerably reduces the quality of the solutions obtained with the Kyriienko-inspired VQAs. It should be noted that the original Kyriienko algorithm has been extensively tested for first-order DEs with very convincing results. It is likely that the proposed Kyriienko-inspired algorithm would require further tuning and possibly a larger number of variational parameters to further improve its performance for high-order DEs. Another key aspect is that the results may highly depend on the initial random values of the variational parameters.

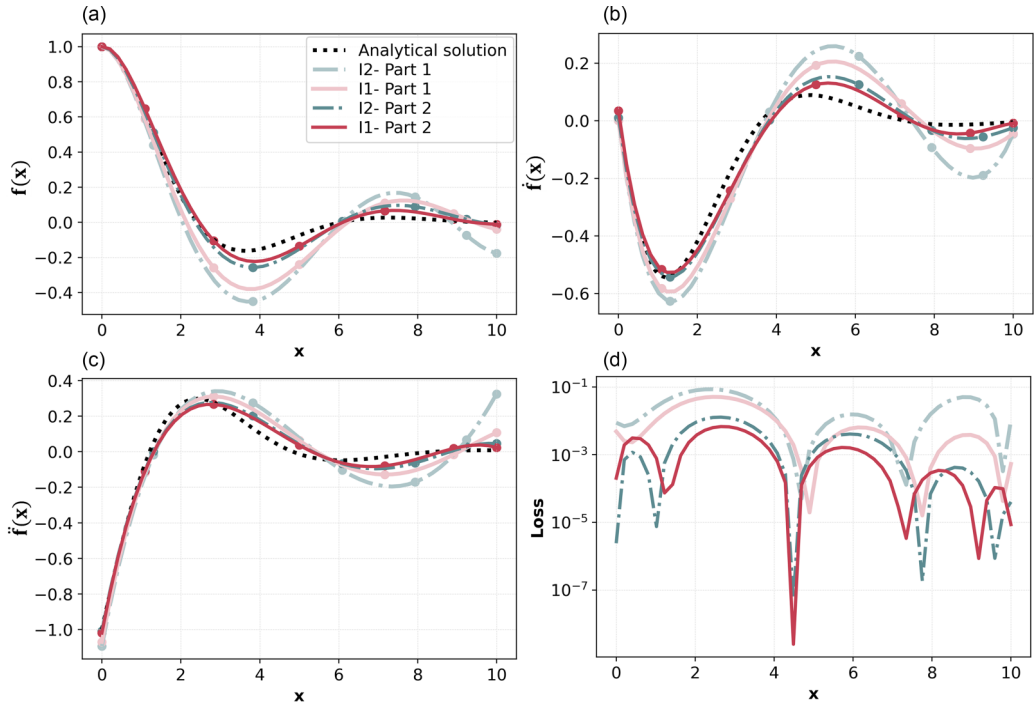

FIG. 11. Results of the new Hadamard-Lagrange VQA for both parts of the optimization process, achieved across five attempts. (a) The function $f$, (b) the first derivative $\dot{f}$, (c) the second derivative $\ddot{f}$, and (d) the DE loss contribution across 50 equispaced points in the interval, plotted on a logarithmic scale. $I1$ stands for the first kind of training set of points while $I2$ stands for the second.

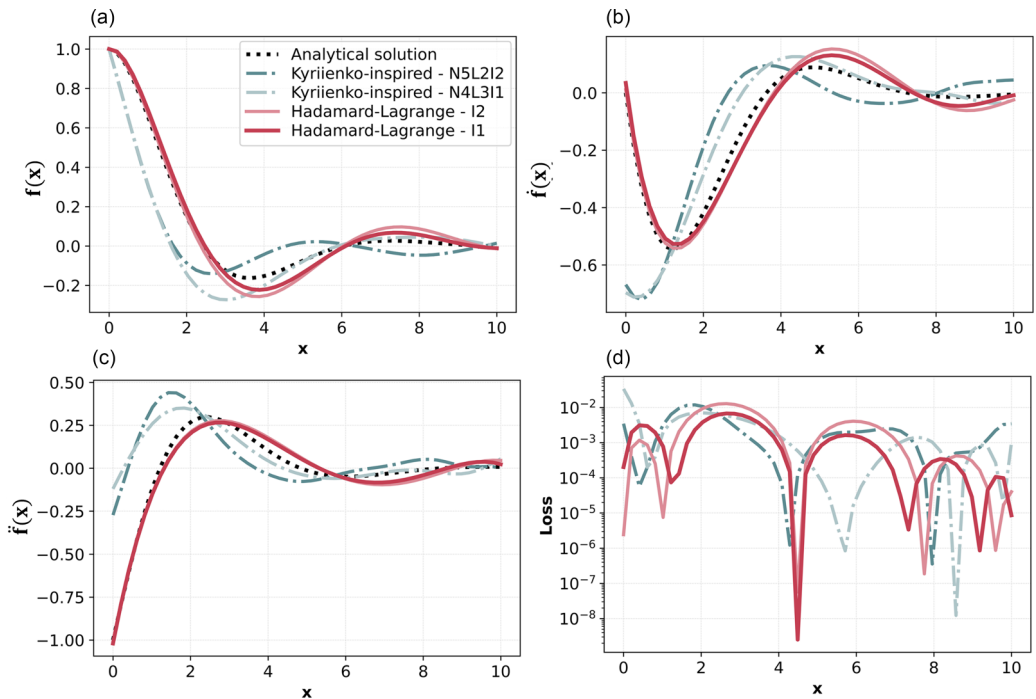

FIG. 12. Results comparison between the new Hadamard-Lagrange VQA and the best outcomes of the Kyriienko-inspired VQA for both kinds of training sets of points, for (a) the function $f$, (b) the first derivative $\dot{f}$, (c) the second derivative $\ddot{f}$, and (d) the DE loss contribution across 50 equispaced points in the interval, plotted on a logarithmic scale.

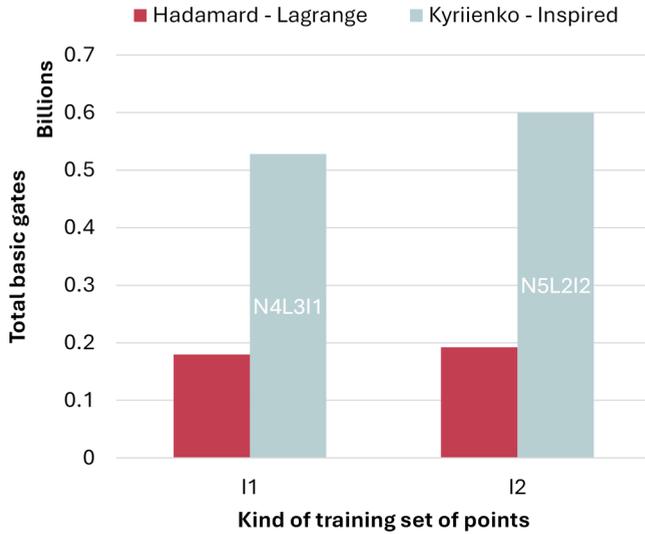

FIG. 13. Gate complexity comparison between the new Hadamard-Lagrange VQA and the best outcomes of the Kyriienko-inspired VQA for both kinds of training sets of points.

A direct comparison of the number of gates required between the solutions obtained with the new VQA and the best outcomes of the Kyriienko-inspired VQA reveals a significant difference, as illustrated in Fig. 13. This comparison positions the new Hadamard-Lagrange algorithms presented in this work as the most precise and efficient for solving, at least, the damped mass-spring system problem (noting that it might be possible to reduce the complexity of the Kyriienko-inspired VQA). Finally, when considering both the quality of the solution and its gate complexity, the optimal solution is derived from the new Hadamard-Lagrange VQA, trained on a set of seven training points of the first kind.

### B. Poisson equation

The Poisson equation is one of the most computationally intensive DEs to solve, requiring very expensive iterative methods. Previous work from Sato *et al.* [31] solved the Poisson equation using a quantum approach for different boundary conditions and a step function $s$ from $1/2^{n/2}$ to $-1/2^{n/2}$ for $n = 5$, using a VQA based on the minimal potential energy, with $n + 1 = 6$ qubits over $2^n = 32$ points. In this study, we solved the same equation using the Hadamard-Lagrange algorithm with a simplified structure on four qubits trained over three points. The Poisson equation can be expressed as

$$\frac{d^2 f}{dx^2} + s = 0, \tag{24}$$

where $f$ is the state field and $s$ the given step function as a source term, plotted in Fig. 14. Different types of boundary conditions can be tested for this equation, namely, periodic, Neumann, and Dirichlet boundary conditions. It is a great opportunity to assess the versatility of the proposed new VQA. This Poisson equation can be analytically solved by splitting the interval in half. The solution over each half interval is a second-degree polynomial where the coefficients are real numbers and depend on the source function and the boundary conditions. The definition of the analytical solutions (depending on the boundary conditions) is further detailed in Appendix D.

#### 1. Hadamard-Lagrange algorithm vs Sato et al. algorithm

The two quantum approaches being compared here for the solution of a Poisson equation are quite different. The Sato *et al.* algorithm uses a discretized approach, where the output is the unitary vector of amplitudes representing the approximated solution function across the specified interval. In contrast, the new algorithm using Lagrange polynomials encoding outputs the approximated solution function, which can be computed for any point or node within the chosen interval. Given its discretized nature, the Sato *et al.* algorithm encodes both the source function $s$ and the algorithm output $\psi_\theta$ into quantum states as vectors of amplitudes [Eq. (5)]. However, this implementation assumes the existence of an efficient unitary operator $\hat{\mathcal{U}}_s$ to prepare the quantum state $|s\rangle = \hat{\mathcal{U}}_s |0\rangle^{\otimes n}$, which may limit its application to other source functions or other DEs.

Our proposed Hadamard-Lagrange algorithm operates with continuous functions. In cases where the source function is discontinuous, as in this particular application, at least the second-order derivative will exhibit a discontinuity. To address this challenge, the algorithm can be applied to sets of subintervals. For example, the algorithm can be initially applied to the first half of the interval, and the solution over the second half can be determined by either symmetry or applying the algorithm again to the second half. Here, the Hadamard-Lagrange algorithm was applied to half of the interval and iterated until it achieved a precision comparable to the Sato *et al.* solution, allowing a fair comparison under similar controlled conditions, as shown in Fig. 14(b). Figure 14(a) illustrates the shape of the source term, with a discontinuity. This figure illustrates how our VQA is eventually converging to the analytical solution. It should also be noted that for our VQA, the DE loss contribution over the half interval is constant, which is not the case for the Sato *et al.* algorithm [dotted line in Fig. 15(b)], for which the loss is much larger at the boundaries of the computational domain. The red line in Fig. 15(b), corresponding to 300 iterations, is the loss obtained with our new VQA, which was stopped when reaching the same averaged-in-space loss as the Sato *et al.* solution. The green line, corresponding to 399, is the loss obtained when reaching the local minimum loss of the Sato *et al.* solution. Note that the data obtained in this section are averaged over five tries to minimize the influence of the random initialization.

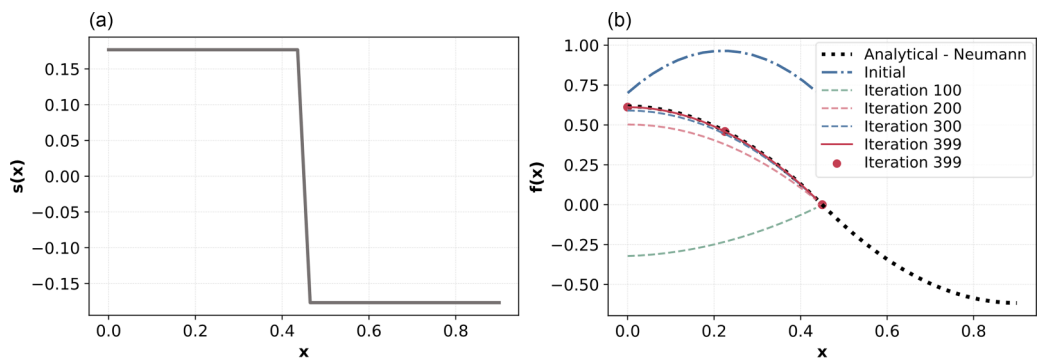

FIG. 14. (a) Source term $s$ used for the Poisson equation application. (b) Iterative evolution of the Hadamard-Lagrange solution for Neumann boundary conditions.

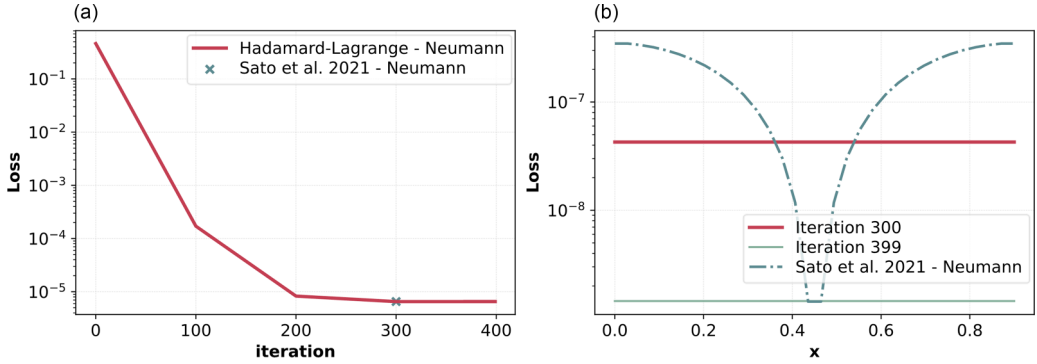

FIG. 15. (a) Iterative evolution of the DE loss $\mathcal{L}_\theta^{\text{DE}}$ obtained with the Hadamard-Lagrange algorithm; equal loss $\mathcal{L}_\theta^{\text{DE}}$ to the Sato *et al.* [31] algorithm is obtained after 300 iterations. (b) Detailed comparison of the DE loss contribution over the interval $\mathcal{L}_{\theta,x}^{\text{DE}}$. Both plots are obtained with Neumann boundary conditions, plotted on a logarithmic scale.

The loss function of the proposed Hadamard-Lagrange algorithm accounts for both the evaluation of the differential equation and the boundary conditions. Figure 16 illustrates the contribution in the total loss function from the differential equation and the constraints of the boundary conditions. It can be seen that the contribution from the boundary conditions constraint is different depending on the type of boundary conditions, with a significant contribution for Neumann boundary conditions and a relatively small contribution for periodic and Dirichlet boundary conditions. This is due to the small difference in the analytical solutions depending on the boundary conditions as explained in Appendix D.

For all boundary conditions (i.e., periodic, Dirichlet, and Neumann), a direct comparison to the analytical solution revealed an overall good agreement for both VQA, even at the boundaries of the computational domain, as seen in Fig. 17. As a reminder, this is achieved with the use of a simplified structure for the circuit of our new VQA. For each boundary condition, the mean absolute error of the solution obtained using the Sato *et al.* algorithm is 4, 3, and 6 times larger than the error obtained for the proposed algorithm, for periodic, Dirichlet, and Neumann boundary conditions, respectively.

To further investigate the performance of both algorithms, it can be helpful to look at the first- and second-order derivatives of the solution of the Poisson equation. Focusing on the case with the Neumann boundary condition, it can be seen in Fig. 18 that the Sato *et al.* solution deviates from the analytical solution at the boundary of the domain for the second-order derivative and near the discontinuity for the first-order derivative. It is important to note that the accuracy of the Sato *et al.* solution could be improved by increasing the spatial discretization, but this would dramatically increase the size of the VQC and consequently the complexity.

### 2. Gate complexity estimation

Similarly to the previous application, the gate complexity of both VQAs has been estimated by the number of circuits as well as the number of basic gates required to complete the calculation. However, this estimation can only be regarded as an order of magnitude estimation due to the fundamentally different approaches of the two VQAs being compared. Moreover, for both algorithms, a different classical optimizer had been used, which significantly impacts the number of iterations and, consequently, the overall gate complexity. As previously mentioned in Sec. II A 3, the BFGS optimizer used by Sato *et al.* [31] has better performances in training linear quantum solvers over few qubits [37] than the Adam optimizer used for the Hadamard-Lagrange algorithm.

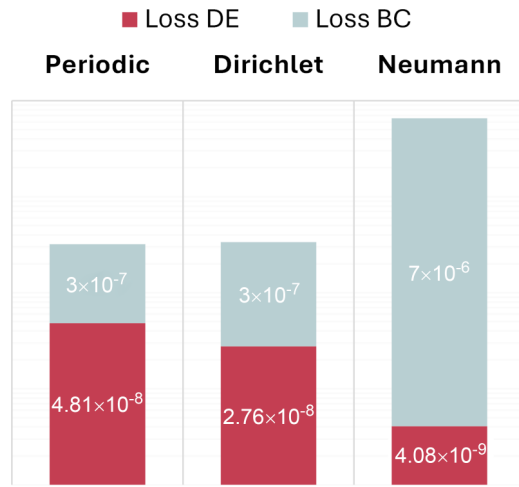

FIG. 16. Overall loss value of the Hadamard-Lagrange algorithm applied to the different boundary conditions, plotted on a logarithmic scale.

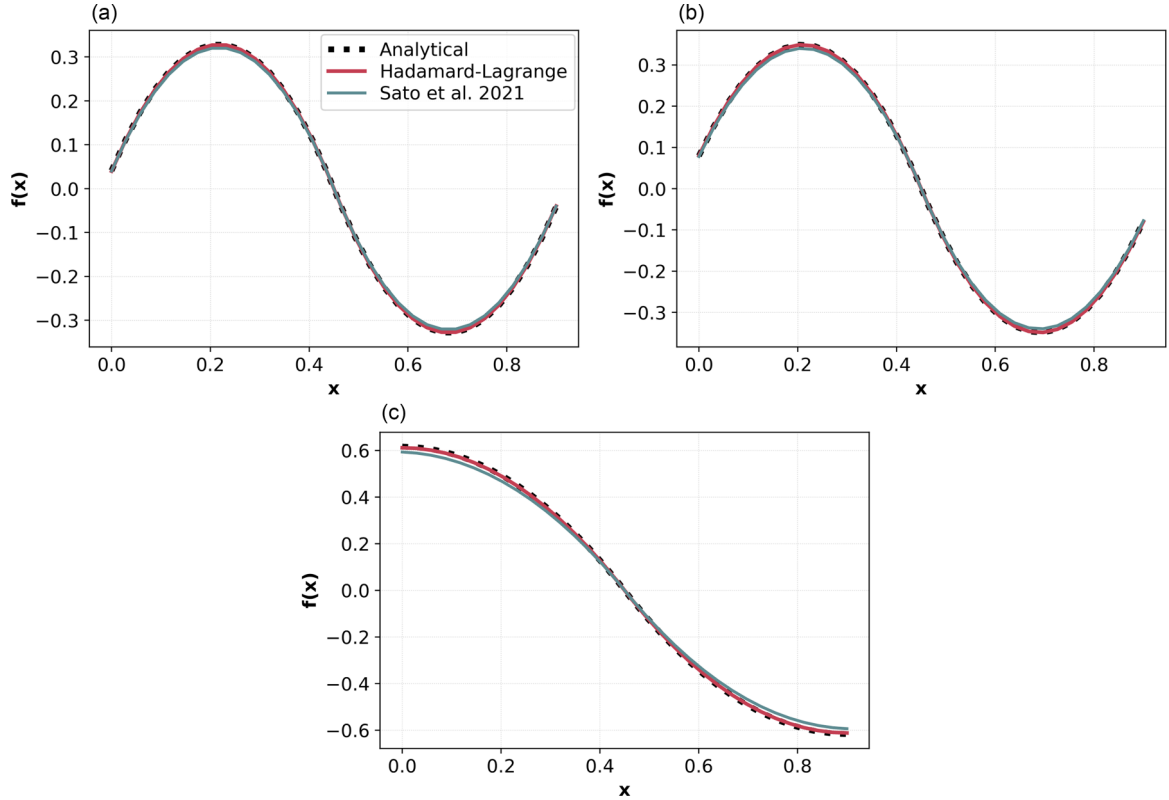

FIG. 17. Comparison of the Hadamard-Lagrange solution with the one obtained with the Sato *et al.* algorithm [31] for (a) periodic, (b) Dirichlet, and (c) Neumann boundary conditions.

For one iteration of the Sato *et al.* algorithm, the main circuit (Fig. 4) is evaluated with 3, 4, or 5 observables for periodic, Dirichlet, and Neumann boundary conditions, respectively, to obtain the loss function and its gradient with respect to the variational vector $\theta$, which contains 45 variational parameters. This requires numerous derivative circuits to compute the gradient and a main circuit with a large variational *Ansatz*. Additionally, the main circuit (Fig. 4) includes an encoding part and, in some cases, a shift part depending on the observable.

In contrast, the proposed Hadamard-Lagrange algorithm requires circuits to evaluate the loss function [Eq. (7)] and its gradient with respect to the variational vector $\theta$ for each iteration. This includes the differential equation loss over the control points $\{x_i\}$ and for the boundary conditions at $x_0$. Here, the variational vector contains only three parameters. This structural difference leads to a significant disparity in gate complexity. Although the Sato *et al.* algorithm [31] requires, on average, half as many iterations to converge to the solution of the Poisson equation, its circuit structure is much more complex, requiring 200 times more gates per iteration compared to the Hadamard-Lagrange algorithm, as illustrated in Fig. 19. This is in line with the observation made for the first application, for which the proposed Hadamard-Lagrange algorithm was already very competitive in terms of gate complexity by comparison to the Kyriienko-inspired algorithm.

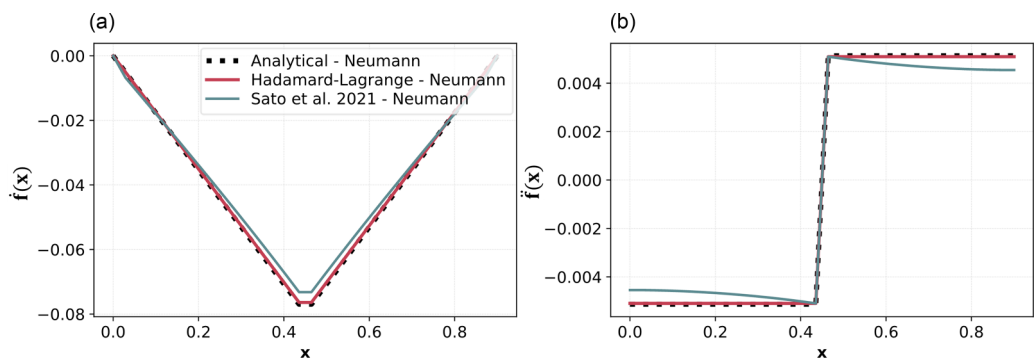

FIG. 18. Comparison of the first- (a) and second-order (b) derivatives of the solution of the Poisson equation for the Neumann boundary condition, obtained with the novel Hadamard-Lagrange algorithm and the Sato *et al.* algorithm [31].

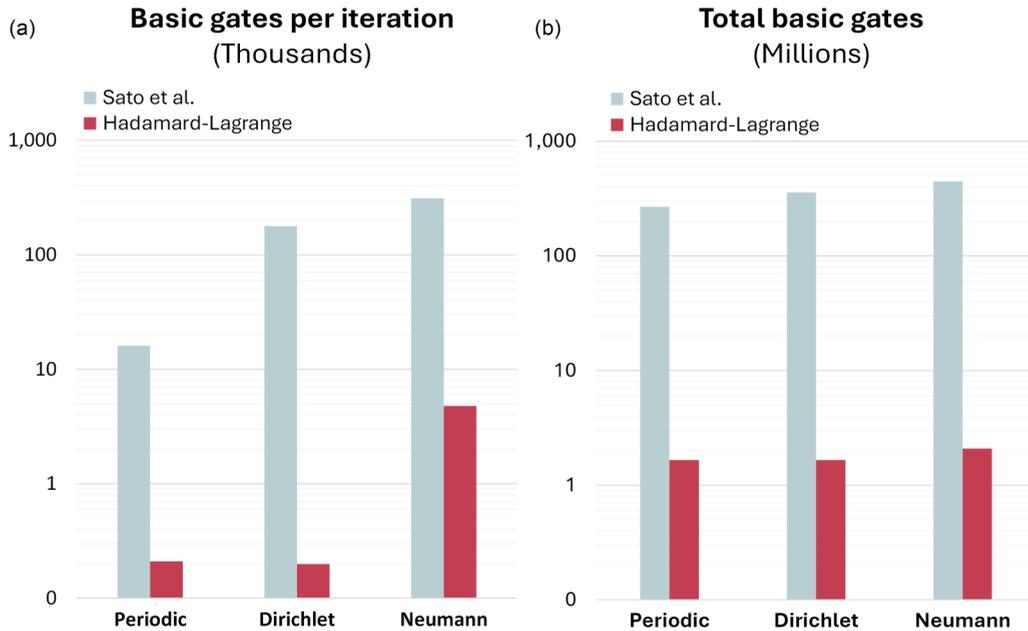

FIG. 19. Gate complexity estimation of the Hadamard-Lagrange algorithm, averaged over five attempts per boundary conditions, compared with the Sato *et al.* algorithm [31], plotted on a logarithmic scale. (a) Basic gates per iteration, (b) total basic gates.

Moreover, while being flexible due to its numerous variational parameters $\theta$, the HEA [54] used in the Sato *et al.* algorithm faces complexities arising from its multilayered structure and the sequence of rotational gates. This large number of parametrized and nonparametrized gates also exacerbates classical optimisation challenges. This complexity may explain why the Sato *et al.* algorithm exhibits a fairly high gate complexity.

## V. CONCLUSIONS AND FUTURE WORK

DEs are critical in various scientific fields such as structural engineering, fluid dynamics, and financial modeling, but can be difficult to solve with traditional computational methods due to their complexity. Recent advancements in quantum computing spur researchers' interest in designing quantum algorithms for solving DEs. This work introduced two distinct architectures of a novel variational quantum algorithm (VQA) based on Lagrange polynomial encoding alongside derivative quantum circuits with Hadamard test differentiation to approximate DE solutions. To demonstrate the potential of this new VQA, it was applied to two well-known DEs: the damped mass-spring system from a given initial condition and the Poisson equation under periodic, Dirichlet, and Neumann boundary conditions. The results showed that the new VQA with its two different circuit architectures has a reduced gate complexity compared to existing variational quantum algorithms while providing similar or better solutions.

Among the algorithms compared in this study, both the Kyriienko-inspired algorithm and the proposed new Hadamard-Lagrange algorithm employ meshless approaches and rely on polynomial fitting, demanding continuity in the definition of the problem. Such feature is not required by the Sato *et al.* algorithm. However, while the latter algorithm needs to discretize the problem and hence can solve only linear or linearized DEs, polynomial fitting algorithms do not require any linearization procedure, which makes them good candidates to solve nonlinear DEs.

Furthermore, these meshless approaches allow to distinguish up to three types of points within the interval $I$: the *training points*, used to evaluate the gradient of the loss function and optimize the variational parameters; the *evaluation points*, used to evaluate the optimized function; and, finally, the *encoding points*, used in the quantum feature map of the Hadamard-Lagrange algorithm. These different sets of points are independent, enabling the training of VQCs on various sets or evolving sets with algorithm iterations. This independence also allows to introduce additional losses at specific points, different from the training set or not, such as boundary conditions.

On the other hand, where the Kyriienko-inspired algorithm encodes the variable $x$ through a nonlinear set of polynomials, which causes some oscillations at the extremity of the interval, encoding the variable with linear polynomials such as with the proposed Hadamard-Lagrange algorithm may not be optimal in an application to a nonlinear problem. In both cases the degree of the fitting polynomial scales with the size of the circuit, which might, at some point, limit their implementation to NISQ devices.

Based on the encouraging results showcased in this study and these comments, we intend to further investigate the potential of our VQA and improve its performance by looking at the following:

(i) *High dimensions and different DEs*. The work presented focused on 1D DEs [damped mass-spring system (DMSS)

and the Poisson equation], but most problems of interest in engineering are in 2D and in 3D, and using a wide range of DEs. We therefore need to assess the performance of our VQA with more than one spatial dimension and for a wide range of DEs (advection-diffusion equation, Burgers's equation, and possibly the Euler equations).

(ii) *Nonlinear DEs.* Nonlinear DEs and their corresponding dynamics are frequently encountered in engineering. Due to the linear evolution of quantum systems, constructing quantum algorithms for solving nonlinear DEs is quite challenging. The next natural step is therefore to look at how our VQA could deal with nonlinear DEs.

(iii) *Tests on quantum computers.* As discussed in the paper determining whether to use the simplified or extended circuit structure for our new VQA depends on the quantum device being employed. The extended circuit structure requires more qubits but fewer gates. Hence, while it reduces noise from gate errors or decoherence, this structure may be prone to errors from qubits or crosstalk. Conversely, the simplified structure shows promise, particularly with the development of efficient mitigation techniques and error gate correction. It would therefore be of interest to investigate the performance of both structures on quantum computers.

*Note added.* Our coauthor, Lorenzo Iannucci, passed away during the preparation of this manuscript.

## ACKNOWLEDGMENTS

The first author would like to acknowledge the Department of Aeronautics, Imperial College London, for supporting this work with a fully funded doctoral studentship. This research was funded by the Engineering and Physical Sciences Research Council in the United Kingdom, Grant No. EP/W032643/1.

## APPENDIX A: HARDWARE-EFFICIENT *ANSATZ* SIMPLIFICATION

The hardware-efficient *Ansatz* [54] was initially described as a quantum circuit comprising layers of a sequence of rotational gates $\hat{R}_Z - \hat{R}_X - \hat{R}_Z$ and CNOT gates, with independent angle parameters $\theta$. However, in the case of a projective measurement onto the $Z$ axis, simplifying this general structure would consist of layers of single rotational gate around the $X$ axis $\hat{R}_X$ and CNOT gates since rotations around the $Z$ axis $\hat{R}_Z$ alone do not affect measurement outcomes in the $Z$ basis.

The magnetization or projective measurement onto the $Z$ axis of a given qubit $|\varphi\rangle$ can be defined as the difference in the probabilities of measuring the qubit in the computational basis states:

$$\begin{aligned}\langle \hat{Z} \rangle &= \langle \varphi | \hat{Z} | \varphi \rangle \\ &= P(|0\rangle) - P(|1\rangle) \\ &= |\langle 0 | \varphi \rangle|^2 - |\langle 1 | \varphi \rangle|^2. \end{aligned} \quad (A1)$$

Since the $\hat{R}_z$ gate only affects the phase factor of the qubit state, the probabilities of measuring the qubit in the computational basis states remain unchanged:

$$|\varphi\rangle = \begin{pmatrix} \alpha \\ \beta \end{pmatrix}; \quad R_z(\phi) = \begin{pmatrix} e^{-i\frac{\phi}{2}} & 0 \\ 0 & e^{i\frac{\phi}{2}} \end{pmatrix}, \quad (A2)$$

$$R_z(\phi)|\varphi\rangle = \begin{pmatrix} \alpha\, e^{-i\frac{\phi}{2}} \\ \beta\, e^{i\frac{\phi}{2}} \end{pmatrix}, \quad (A3)$$

$$\begin{aligned}\langle\varphi|R_Z(\phi)^\dagger \hat{Z} R_Z(\phi)|\varphi\rangle &= |\langle 0|R_Z(\phi)\varphi\rangle|^2 - |\langle 1|R_Z(\phi)\varphi\rangle|^2 \\ &= |\alpha|^2 - |\beta|^2 \\ &= |\langle 0|\varphi\rangle|^2 - |\langle 1|\varphi\rangle|^2 \\ &= \langle\varphi|\hat{Z}|\varphi\rangle. \end{aligned} \quad (A4)$$

## APPENDIX B: LAGRANGE QUANTUM FEATURE MAP

For both circuit structures presented in this paper, the first register is evaluated via the same cost operator defined in Eq. (19) to obtain a Lagrange interpolating polynomial given in Eq. (20).

### 1. Inspiration from the Hadamard test structure

The Lagrange polynomial encoding employed in both circuit structures is inspired by the Hadamard test, illustrated in Fig. 2(a). This approach is particularly advantageous due to its differentiation techniques that require only a minimal number of circuit evaluations. In the classical Hadamard test, the expectation value of the Pauli-Z operator measured on the control qubit corresponds to $\text{Re}(\langle 0|\hat{\mathcal{U}}|0\rangle)$. To adapt this principle for Lagrange encoding, the controlled unitary is replaced by a network of CNOT and RY gates, allowing an efficient encoding of the Lagrange basis while avoiding redundancy of gates. The underlying structure of the encoding can be made explicit by rewriting the Hadamard test as follows:

$$|q_0, a\rangle = (H \otimes I)\text{CNOT}_{q_0,a}(I \otimes \hat{R}_Y[\varphi_i(x)])\text{CNOT}_{q_0,a}(H \otimes I), \quad (B1)$$

where $\text{CNOT}_{q_0,a}$ denotes a controlled-NOT gate with control on the first qubit and target on the ancilla qubit. This circuit first prepares a Bell state and then introduces the interpolation point $x_i$ through the function $\varphi_i = \arccos[(x - x_i)/2]$. The resulting quantum state is

$$|q_0, a\rangle = \cos(\varphi_i/2)|00\rangle + \sin(\varphi_i/2)|11\rangle. \quad (B2)$$

A measurement of the Pauli-Z observable on the first qubit then yields

$$\begin{aligned}\langle q_0|\hat{Z}|q_0\rangle &= [\cos(\varphi_i/2)]^2 - [\cos(\varphi_i/2)]^2 \\ &= \cos(\varphi_i) = (x - x_i)/2. \end{aligned} \quad (B3)$$

### 2. Lagrange basis polynomials encoding

This procedure is extended across $n$ qubits and following the entanglement scheme of one of the two structures presented in the Methods section, the quantum feature map encodes on each qubit in the first register, the following

polynomial :

$$\langle f_{\varphi,0}(x)|\hat{Z}_j|f_{\varphi,0}(x)\rangle = \frac{1}{2^{n-1}} \prod_{i\neq j}^{n} (x-x_i), \quad \text{(B4)}$$

which is proportional to the numerator of the Lagrange basis polynomials $L_j$:

$$L_j(x) = \prod_{i\neq j}^{n} \frac{(x-x_i)}{(x_j-x_i)}. \quad \text{(B5)}$$

To recover the full Lagrange basis polynomials, the denominator is obtained by evaluating the quantum feature map at the interpolation node $x = x_j$, yielding a constant denoted by $\rho_j$ as defined in Eq. (19). Notably, $\rho_j$ is independent of the variational parameters and can therefore be precomputed and stored for reuse, assuming storage resources are available. This constant also compensates for the prefactor $1/2^{n-1}$ introduced by the measurement process, ensuring that the encoded polynomial amplitudes are correctly scaled.

### 3. Lagrange interpolation polynomial

The final Lagrange interpolation polynomial is assembled by applying a variational *Ansatz* on top of the Lagrange basis states. In this formulation, trainable weights are introduced via the variational parameters $\theta$:

$$\langle f_{\varphi,\theta}(x)|\hat{Z}_j|f_{\varphi,\theta}(x)\rangle = \frac{1}{2^{n-1}} \prod_{i\neq j}^{n} \alpha_{i,j}(\theta)(x-x_i). \quad \text{(B6)}$$

Summing the contributions of all qubits in the first register, the quantum circuit evaluates the complete interpolated function:

$$f(x) = \sum_{j=1}^{n} \frac{\langle f_{\varphi,\theta}(x)|\hat{Z}_j|f_{\varphi,\theta}(x)\rangle}{\langle f_{\varphi,0}(x_j)|\hat{Z}_j|f_{\varphi,0}(x_j)\rangle} = \sum_{j=1}^{n} \alpha_j(\theta) \prod_{i\neq j}^{n} \frac{(x-x_i)}{(x_j-x_i)}$$

$$= \sum_{j=1}^{n} \alpha_j(\theta) L_j(x). \quad \text{(B7)}$$

## APPENDIX C: ESTIMATION OF THE GATE COMPLEXITY

In this Appendix, the methodology used to determine the number of circuits and basic quantum gates required for one iteration is detailed. The basic gates considered are Pauli and Clifford gates.

### 1. Kyriienko-inspired algorithm

Regarding the quantum solver inspired by the work of Kyriienko *et al.* [29], one iteration involves, as detailed in the Background and Methods sections, circuits for the functions $f_\theta$, its derivatives $\frac{df_\theta}{dx}, \frac{d^2 f_\theta}{dx^2}$, as well as their respective gradients $\nabla_\theta(f_\theta), \nabla_\theta(\frac{df_\theta}{dx}), \nabla_\theta(\frac{d^2 f_\theta}{dx^2})$ for all controlled points and boundary conditions. In this algorithm, the gradients are obtained using the parameter shift rule method, which requires two circuits for a partial derivative. As a result, the number of circuits per iteration is defined as follows:

$$N_{\text{circuits/iteration}}$$
$$= N_{\text{points}}(1 + 2N_{\text{parameters}})$$
$$\times \left[ N_{\text{circuits}}(f_\theta) + N_{\text{circuits}}\left(\frac{df_\theta}{dx}\right) + N_{\text{circuits}}\left(\frac{d^2 f_\theta}{dx^2}\right) \right], \quad \text{(C1)}$$

where $N_{\text{circuit}}$ is zero if the respective component does not appear in the differential equation. In the case of the damped-mass-spring-system problem solved in Sec. IV A, the differential equation is composed by all the components mentioned above.

$$N_{\text{points}} = N_{\text{controlled points}} + 1,$$
$$N_{\text{parameters}} = N_{\text{qubits}} N_{\text{layers}},$$
$$N_{\text{circuits}}(f_\theta) = 1,$$
$$N_{\text{circuits}}\left(\frac{df_\theta}{dx}\right) = 2N_{\text{qubits}},$$
$$N_{\text{circuits}}\left(\frac{d^2 f_\theta}{dx^2}\right) = 4N_{\text{qubits}}^2. \quad \text{(C2)}$$

The circuits are constructed across five qubits with a variational *Ansatz* depth of 2. Moreover, the algorithm employs a quantum circuit structure based on basic quantum gates such as CNOT, *RY*, and *RX* quantum gates. The count of gates per circuit can be expressed as follows:

$$N_{\text{basic gates/circuit}} = N_{\text{qubits}} + 2N_{\text{parameters}}. \quad \text{(C3)}$$

### 2. Hadamard-Lagrange algorithm

As outlined in the Methods section and depicted in Fig. 5, each iteration of the Hadamard-Lagrange algorithm draws inspiration from the methodology proposed by Kyriienko *et al.* [29]. Consequently, the number of circuits per iteration can be defined by Eq. (C1). However, due to variations in encoding and structure, the derivative circuits with respect of $x$ are obtained via the Hadamard test differentiation method, necessitating two times less circuits per derivative.

$$N_{\text{circuits}}\left(\frac{df_\theta}{dx}\right) = N_{\text{interpolation points}},$$
$$N_{\text{circuits}}\left(\frac{d^2 f_\theta}{dx^2}\right) = N_{\text{interpolation points}}^2. \quad \text{(C4)}$$

In both the extended and simplified structures, the derivative circuit necessitates an additional gate. Notably, the reduced structure exhibits a slightly higher count of gates overall, despite being composed of half the number of qubits.

For the extended structure,

$$N_{\text{basic gates/circuit}}(f_\theta) = 5N_{\text{interpolation points}} + 2N_{\text{parameters}} \quad \text{(C5)}$$

For the simplified structure,

$$N_{\text{basic gates/circuit}}(f_\theta) = 5N_{\text{interpolation points}} + 2N_{\text{parameters}}$$
$$+ \lfloor N_{\text{interpolation points}} /2 \rfloor. \quad \text{(C6)}$$

And in both cases,

$$N_{\text{basic gates/circuit}}\left(\frac{df_\theta}{dx}\right) = N_{\text{basic gates/circuit}}(f_\theta) + 1,$$

$$N_{\text{basic gates/circuit}}\left(\frac{d^2 f_\theta}{dx^2}\right) = N_{\text{basic gates/circuit}}(f_\theta) + 2. \quad \text{(C7)}$$

Additionally, in the application of this algorithm, the number of qubits can evolve throughout the algorithm depending on the number of interpolation points. Consequently, the estimation of the number of circuits as well as the number of gates should be considered step by step.

### 3. Sato *et al.* Algorithm

The Sato *et al.* algorithm [31] was designed to solve the Poisson equation example described in the Results section. This algorithm employs a fundamentally different approach and circuit structure. For each iteration, evaluating the loss function, defined as the total potential energy of the system requires 3, 4, or 5 circuits for periodic, Dirichlet, or Neumann boundary conditions, respectively. Additionally, the derivative of the loss function for each variational parameter must be evaluated. This is achieved using the Hadamard test differentiation method, which requires only one modified circuit per partial derivative. Thus, the number of circuits needed for one iteration can be expressed as follows:

$$N_{\text{circuits/iteration}} = N_{\text{observables}} \left(1 + N_{\text{parameters}}\right), \quad \text{(C8)}$$

where $N_{\text{observables}} = 3$, 4, or 5 depending on the boundary condition, and $N_{\text{parameters}} = N_{\text{layers}} \times N_{\text{parameters/layer}} + N_{\text{encoding qubits}}$. For the Poisson equation solved in the Results section, five qubits are used for encoding, the number of layers of the *Ansatz* is set to 5, and the number of parameters per layer is 8, resulting in a total of 45 parameters. To account for the gate complexity of the algorithm, the number of basic quantum gates (Pauli and Clifford gates) per iteration is determined as follows:

$$N_{\text{basic gates/circuit}} = 3 + N_{\text{basic gates}}(\hat{\mathcal{U}}_b) + N_{\text{basic gates}}(\hat{\mathcal{U}}_\theta)$$
$$+ N_{\text{shift circuits}} N_{\text{basic gates}}(\hat{\mathcal{U}}_{\text{shift}}), \quad \text{(C9)}$$

with

$$N_{\text{basic gates}}(\hat{\mathcal{U}}_b) = N_{\text{encoding qubits}} + 1,$$
$$N_{\text{basic gates}}(\hat{\mathcal{U}}_{\text{shift}}) = 1 + N_{\text{encoding qubits}} N_{\text{basic gates}}(\text{MCP}),$$
$$N_{\text{basic gates}}(\hat{\mathcal{U}}_\theta) = N_{\text{parameters}} N_{\text{basic gates}}(\text{CRY})$$
$$+ N_{\text{layers}} N_{\text{basic gates}}(\text{MCP}). \quad \text{(C10)}$$

The CRY quantum gate refers to the controlled rotational gate around the $Y$ axis, which can be decomposed into four basic quantum gates. The MCP quantum gate, or the multicontrolled phase gate, is equivalent to a Toffoli quantum gate in the case of two controlled qubits, which can be decomposed into 18 basic quantum gates.

## APPENDIX D: ANALYTICAL SOLUTIONS

### 1. Damped mass-spring system

In this study, we investigate the dynamics of a damped mass-spring system described by the second-order differential equation, referenced as Eq. (21). The corresponding characteristic equation is formulated as

$$m\lambda^2 + b\lambda + k = 0. \quad \text{(D1)}$$

For the scenario explored in this paper, the discriminant $\Delta = b^2 - 4mk$ is negative, indicating an underdamped case. Consequently, the solution takes the form

$$f(x) = e^{\alpha x}[C_1 \cos(\beta x) + C_2 \sin(\beta x)], \quad \text{(D2)}$$

where $\alpha$ and $\beta$ are the real and imaginary parts of the complex-conjugate roots $\lambda_{1,2} = \alpha \pm i\beta$. The coefficients $C_1$ and $C_2$ are determined by the initial conditions $f(x_0) = u_0$ and $f'(x_0) = \dot{u}_0$.

### 2. Poisson equation

The Poisson equation is defined in Eq. (24), with a given source term defined as follows:

$$s(x) = \begin{cases} +(1/2)^{n/2} & \text{if } x < \frac{a+b}{2}, \\ -(1/2)^{n/2} & \text{if } x > \frac{a+b}{2}, \end{cases} \quad \text{(D3)}$$

with $n = 5$, over the interval $[a, b]$. In the work of Sato *et al.* [31], this Poisson equation is solved over 32 nodes, defining the interval as $[\![0, 31]\!]$. As mentionned in the Results section, the analytical solution can be determined by splitting the interval in half. The solution over each half is a second-degree polynomial as expressed in the following equation:

$$f(x) = \begin{cases} Ax^2 + B_1 x + C_1 & \text{if } x < \frac{a+b}{2}, \\ Ax^2 + B_2 x + C_2 & \text{if } x > \frac{a+b}{2}, \end{cases} \quad \text{(D4)}$$

with $A = -\frac{1}{2}^{\frac{n}{2}+1}$. $B_i$ and $C_i$ are real coefficients which depend on the boundary conditions. In every case, the continuity of the middle point $x_m = \frac{a+b}{2}$ for the solution function and its first derivative gives the following definitions:

$$\lim_{x \to x_m^+} f(x) = \lim_{x \to x_m^-} f(x) = f(x_m), \quad \text{(D5)}$$

$$\lim_{x \to x_m^+} \frac{df}{dx}(x) = \lim_{x \to x_m^-} \frac{df}{dx}(x) = \frac{df}{dx}(x_m). \quad \text{(D6)}$$

Other conditions depend on the boundary. For the periodic boundary condition, the solution follows $f(x + T) = f(x)$, with the period $T$ defined in Sato *et al.* [31] work as $T = b - a + 1 = 32$. The Dirichlet boundary condition consists in defining the values at the extremities of the interval. In Sato *et al.* [31] work, it is defined as $f(a - 1) = f(b + 1) = 0$. Finally, the Neumann boundary condition indicates the values of the first-order derivative at the extremities, which is defined here as $f'(a) = f'(b) = 0$. All considered, the analytical solutions for each boundary equation are

(i) Periodic:

$$f_P(x) = \begin{cases} A\left[x^2 - \frac{1}{2}(3a+b-1)x + \frac{1}{2}\left(a-\frac{1}{2}\right)(a+b+1)\right] & \text{if } x < x_m, \\ 0 & \text{if } x = x_m, \\ -A\left[x^2 - \frac{1}{2}(3b+a+1)x + \frac{1}{2}\left(b+\frac{1}{2}\right)(a+b+1)\right] & \text{if } x < x_m, \end{cases} \quad \text{(D7)}$$

(ii) Dirichlet:

$$f_D(x) = \begin{cases} A\left[x^2 - \frac{1}{2}(3a+b-2)x + \frac{1}{2}(a-1)(a+b)\right] & \text{if } x < x_m, \\ 0 & \text{if } x = x_m, \\ -A\left[x^2 - \frac{1}{2}(a+3b+2)x + \frac{1}{2}(b+1)(a+b)\right] & \text{if } x < x_m, \end{cases} \quad \text{(D8)}$$

(iii) Neumann:

$$f_D(x) = \begin{cases} A\left[x^2 - 2ax + \frac{1}{4}(3a^2 + 2ab - b^2)\right] & \text{if } x < x_m, \\ 0 & \text{if } x = x_m, \\ -A\left[x^2 - 2bx + \frac{1}{4}(3b^2 + 2ab - a^2)\right] & \text{if } x < x_m. \end{cases} \quad \text{(D9)}$$